\documentclass[useAMS,usenatbib]{mn2e}
\usepackage[cp1251]{inputenc}
\usepackage{amsmath}
\usepackage{amssymb}
\usepackage{euscript}
\usepackage[dvips]{graphicx}
\usepackage{epsfig}
\usepackage[usenames,dvipsnames]{color}
\usepackage{longtable}
\usepackage{ulem}

{}

\title[H$_2$ absorption system in SDSS]
{Molecular hydrogen absorption systems in Sloan Digital Sky Survey}

\author[Balashev et al.]
{S.A.~Balashev$^{1,2}$\thanks{E-mail: balashev@astro.ioffe.ru},
V.V.~Klimenko$^{1,2}$, A.V.~Ivanchik$^{1,2}$, D.A.~Varshalovich$^{1,2}$, 
\newauthor P.~Petitjean$^3$, P. Noterdaeme$^3$\\
\vspace{4pt}\\
$^1$Ioffe Physical-Technical Institute of RAS, {Polyteknicheskaya 26}, 194021 Saint-Petersburg, Russia\\
$^2$St.-Petersburg State Polytechnical University, {Polyteknicheskaya 29}, 195251 Saint-Petersburg, Russia \\
$^3$Institut d'Astrophysique de Paris, UPMC-CNRS, UMR7095, 98bis bd Arago, 75014
Paris, France \\
}
\begin{document}

\date{Accepted 7 February 2014. Received 31 August 2013}

\pagerange{\pageref{firstpage}--\pageref{lastpage}} \pubyear{2009}

\maketitle

\label{firstpage}

\begin{abstract}
We present a systematic search for molecular hydrogen absorption systems at high redshift in quasar spectra from the 
Sloan Digital Sky Survey (SDSS) II Data Release 7 and SDSS-III Data Release 9. We have selected candidates using a modified profile fitting technique
taking into account that 
the Ly$\alpha$ forest can effectively mimic H$_2$ absorption systems at the resolution of SDSS data. 
To estimate the confidence level 
of the detections, we use two methods: a Monte-Carlo sampling and an analysis of control samples. 
The analysis of control samples allows us 
to define regions of the spectral quality parameter space where H$_2$ absorption systems can be confidently identified. 
We find that H2 absorption systems with column densities $\log {\rm N_{H_2}} > 19$ can be detected in only less than 3\% of SDSS quasar spectra.
We estimate the upper limit on the detection rate of saturated H$_2$ absorption 
systems ($\log {\rm N_{H_2}} > 19$) in Damped Ly-$\alpha$ (DLA) systems to be about 7\%. We provide a sample of 23 confident 
H$_2$ absorption system candidates that would be interesting to follow up with high resolution spectrographs. There is a 
1$\sigma$\,\,$r-i$ color excess and non-significant $A_{\rm V}$ extinction excess in quasar spectra with an H$_2$ candidate compared to standard DLA-bearing quasar spectra.  
The equivalent widths (EWs) of C\,{\sc ii}, Si\,{\sc ii} and Al\,{\sc iii} (but not Fe\,{\sc ii}) absorptions associated with H$_2$ candidate DLAs are larger compared to standard DLAs. This is probably related to a larger spread in velocity of the absorption lines in the H$_2$ bearing sample. 
\end{abstract}


\section{Introduction}
\label{introduction}
\noindent
The Sloan Digital Sky Survey (SDSS; \citealt{York2000}) is one of the largest optical surveys of modern astrophysics. 
One of the major goals of this survey is to study  large scale structures in the nearby Universe $z<0.7$, from the spatial distribution of 
galaxies \citep{York2000}. The survey targets millions of objects (galaxies, stars and quasars) 
by imaging and spectroscopy in the optical wavelength band. The recently published ninth data release 9 (DR9) \citep{Ahn2012} 
contains over 80\,000 spectra of high redshift quasars \citep{Paris2012}. 

Absorption lines in quasar spectra allow one to study the intergalactic and interstellar matter located along the line of sight to 
the quasar. Quasars are routinely detected up to $z\sim6$, which corresponds to more than 12 Gyr ago. The Baryon Oscillation Spectroscopic 
Survey (BOSS; Schlegel et al. 2007; Dawson et al. 2012), one of the main projects of  SDSS-III (the third generation of SDSS) is 
primarily focused on the analysis of the spatial distribution of luminous red galaxies (LRGs) and absorptions in the Ly$\alpha$ forest. 
The later enables to determine the baryon acoustic oscillation (BAO) scale at z\,$\sim 2.5$ that corresponds to an epoch before 
dark energy dominates the expansion of the Universe \citep{Busca2013}. 
In addition to this primary goal, BOSS spectra will allow one to study of broad absorption line (BAL) systems arising in the vicinity of  
Active Galaxy Nucleus (AGN), as well as intervening metal absorption line systems (such as C~{\sc iv} and Mg~{\sc ii}) associated 
with clouds located into the halo of intervening galaxies \citep{Quider2011, Zhu2013}.

Damped Lyman Alpha systems (DLAs) are identified by a broad H~{\sc i} Ly-$\alpha$ absorption line with prominent 
saturated Lorentz wings. Statistical analysis shows that DLA systems are the main reservoir of neutral gas at high redshifts 
\citep{Prochaska2009, Noterdaeme2009}. It is believed that these systems are associated with disks of galaxies or their close 
vicinity with impact parameter less than 20 kpc \citep{Fynbo2011, Krogager2012}. 
Numerous species are seen in DLAs (e.g. C\,{\sc i} to C\,{\sc iv} and Si\,{\sc i} to  Si\,{\sc iv}) and the typical velocity spread of metal absorption 
lines is about 100--500 km/s.
All this makes DLA systems relatively easy to detect at intermediate resolution. About 12,000 DLA/sub-DLA candidates with log N(H\,{\sc i})$>=$20 were detected in SDSS-DR9 
\citep{Noterdaeme2012}. It has been shown that a small fraction of DLAs hosts molecular hydrogen \citep{Noterdaeme2008} which corresponds to diffuse and translucent neutral clouds embedded in warm neutral interstellar medium.
In some cases HD molecules are detected \citep{Varshalovich2001, Noterdaeme2008b, Balashev2010} as well as CO molecules 
\citep{Srianand2008, Noterdaeme2011}. 
Observations of molecular hydrogen clouds at the high redshift Universe provide a unique opportunity to study several issues. 
Molecular hydrogen is believed to be an indicator of the cold phase of the interstellar medium which is the raw material 
for the star-formation -- it was found that regions with high H$_2$ abundance are correlated with the star-formation regions 
\citep{Krumholz2012}. The measurement of relative abundances of different H$_2$ rotational levels allows one to determine the physical 
conditions in this cold interstellar medium --  kinetic temperature, UV radiation field and possibly number density. There are several cosmological 
problems related to high redshift molecular hydrogen absorption systems: (i) the detection of HD gives the possibility of a complementary 
approach to the determination of the primordial deuterium abundance \citep{Ivanchik2010,Balashev2010}; (ii) constraints on the possible variation of the proton 
to electron mass ratio can be obtained (e.g. \citealt{Thompson1975, Ivanchik2005, Wendt2012, Rahmani2013}); (iii) the CMBR temperature can be 
measured at high redshift from the populations of the CO rotational levels (e.g. \citealt{Noterdaeme2011}) and  C\,{\sc i} fine-structure 
levels (e.g. \citealt{Songaila1994, Srianand2000}).

Since the first detection by \citet{Levshakov1985}, about 20 H$_2$ absorption systems have been detected at high redshifts ($z>1$, see \citealt{Ge1997, Ledoux2002, Ledoux2003, Reimers2003, Cui2005, Noterdaeme2008, Srianand2008, Jorgenson2009, Jorgenson2010, Malec2010, Noterdaeme2010, Fynbo2011, Guimaraes2012}). Additionally, two 
systems were detected in spectra of GRB afterglows \citep{Prochaska2009b, Kruhler2013}, one system was detected at intermediate redshift $z\sim 0.5$ \citep{Crighton2013}, and four systems \citep{Noterdaeme2009, Noterdaeme2011} were detected by means of CO molecules (H$_2$ and HD transitions in these systems are redshifted out of the observed wavelength range). Most of the detection of molecular hydrogen absorption systems 
have been performed with high signal to noise ratio (S/N) and high resolution spectra with typically 3 hours exposures on a 8~m class telescope.
On the other hand, there are about 80,000 high redshift quasar spectra obtained in the course of SDSS to be search for H$_2$. 
The main disadvantages of SDSS spectra for detection of molecular hydrogen systems are that they have intermediate spectral 
resolution $R$$\sim$2000 and usually low S/N ($<$4 over the wavelength range where H$_2$ absorption lines are to be found). 

In this paper we show however that it is possible to find H$_2$-bearing systems in SDSS spectra provided that the column density is large
enough. Although such systems are  rare, the large number of SDSS quasar spectra makes their detection possible. 
We provide a list of H$_2$ candidates for follow-up with high resolution observations.
The paper is organized as follows. In Section~2 we give a brief description 
of the data. Section~3 describes a searching criterion to be used. 
We define a quantitative estimate of the confidence level or false identification probability in Section~4. 
The final list of the H$_2$ system candidates is given in Section~5 and we 
investigate some properties 
of this sample in Section~6 before conclusions are drawn in Section~7.

\section{Data}
\label{data}
\begin{figure*}
\begin{center}
        \includegraphics[width=1.0\textwidth]{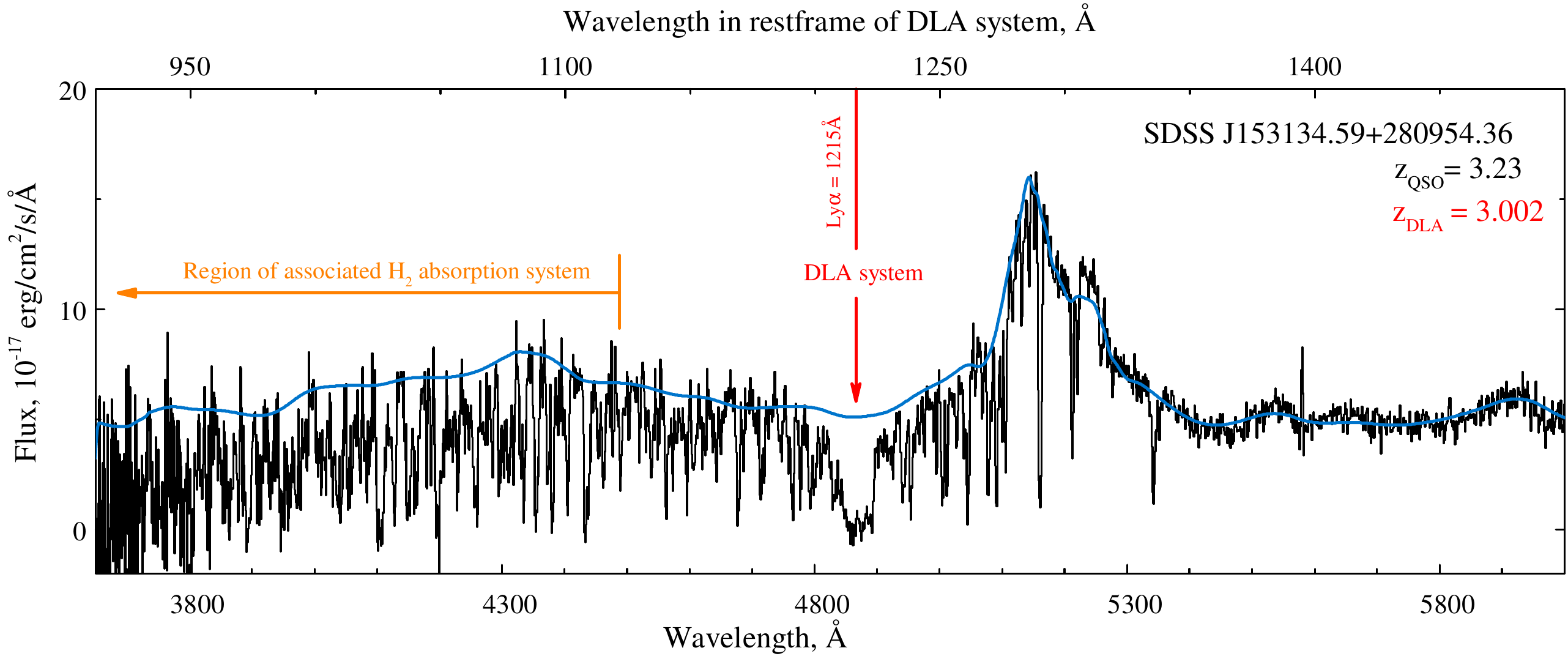}
        \caption{\rm SDSS spectrum of quasar J153134.59+280954.36 ($z_{\rm QSO}=3.23$). The top axis shows the wavelength scale
in the restframe of the DLA system which was identified at $z_{\rm DLA}=3.002$. The orange line marks the region of the spectrum where H$_2$ absorptions (associated with the DLA system) are located. Note, that the signal-to-noise ratio in this region is about 10, 
which is well above the median S/N of the SDSS spectra. The blue curve shows the estimated quasar unabsorbed continuum.}
        \label{SDSSspectrum}
\end{center}
\end{figure*}

\noindent
We used the spectroscopic data from the SDSS.
The Ninth Data Release 9 (DR9; \citealt{Ahn2012}) presents the first spectroscopic data from the Baryon Oscillation Spectroscopic Survey (BOSS; Schlegel et al. 2007; Dawson et al. 2012) and contains 87\,822 primarily high redshifted quasars (78\,086 are new detections) detected over an area of 3\,275 deg$^2$. The data from previous parts of projects SDSS-I and SDSS-II were presented in the Seventh Data Release (DR7; \citealt{Abazajian2009}) and Quasar Catalogs \citep{Schneider2010} include 105\,783 quasars in an area of 9\,380 deg$^2$. 
Since SDSS-III preferentially targets high redshift quasars, most of the searched spectra are from DR9.
In addition, we found that the improved quality of the BOSS spectra is an important factor for the detection of H$_2$ absorption systems.
However, we also applied our searching routine to the DR7 catalog. The spectrum of J153134.59+280954.36 is shown in Fig.~\ref{SDSSspectrum} as an example.

It is commonly accepted that H$_2$ molecular clouds are related to the neutral medium. Hence high column density ($\log$$N$$>16$, $N$ in cm$^{-2}$) H$_2$ absorption systems have to be associated with large amount of neutral hydrogen, i.e. to be associated with DLA systems. 
Since the Ly-$\alpha$ transition being at 1215.67\AA, DLA systems in SDSS can be identified only for redshifts $z\gtrsim2.15$ and z$\gtrsim2$ for 
DR7 and DR9, respectively. 
This gives 14\,616 quasar spectra from SDSS-DR7 and 61\,931 from BOSS SDSS-DR9. In these spectra 1\,426 and 12\,068 DLA systems were 
detected in DR7 \citep{Noterdaeme2009} and DR9 \citep{Noterdaeme2012}, respectively. Note that a part of high redshift DR7 spectra was 
reobserved by BOSS therefore some fraction of quasars is common to both DR7 and DR9.

We used only quasar spectra for which at least one H$_2$ absorption line associated with corresponding DLA system falls in the SDSS wavelength range. The blue limits of the SDSS-II and BOSS spectrographs are 3800 \AA\,\, and about 3570 \AA, respectively. Molecular hydrogen lines from J=0,1 levels are located in the wavelength range 912-1110 \AA\, in restframe (see Fig.~\ref{SDSSspectrum}). It restricts redshifts of suitable DLA systems to values z$_{DLA}>2.42$ and z$_{DLA}>2.22$ from DR7 and DR9, respectively. About 1200 and 10\,000 quasar spectra satisfy this criteria from DR7 and DR9, respectively. These quasars make up the sample (which we refer below as S$_{\rm DLA}$) to search for H$_2$ absorption systems. 

We additionally have built the sample of non-BAL and non-DLA quasar spectra from SDSS DR9 catalog which satisfied our QSO redshift and S/N conditions
($z_{\rm QSO} > 2.22$ and S/N$>2$).  
We denote this sample as S$_{\rm nonDLA}$. It contains about 40\,000 quasar spectra. In principle, we can expect that spectra from this sample contain no
or very few\footnote{Although the completeness of the DLA sample is not unity, in particular at the low N(H\,{\sc i}) end, the
overwhelming number of spectra will not contain H$_2$ system} H$_2$ absorption systems because no corresponding DLA
systems were found. This sample will be used as a control sample to test the searching routine and to estimate the detection 
limit of H$_2$ absorption systems.

\section{Search of H$_2$ absorption systems}
\label{procedure}
\noindent

With such a large number of spectra we need an automatic procedure to search for H$_2$. 
For this purpose we used a modification of the standard profile fitting technique which compares an observed quasar spectrum with 
a synthetic H$_2$ absorption spectrum. 

\subsection{Continuum determination}
\label{continuum}
\noindent
The determination of the quasar continuum is the first important step of any profile fitting routine.
We used a combination of two methods: the principal component analysis (PCA) and the iterative smoothing. 

The first method reproduces an unabsorbed flux over the Ly$\alpha$ forest by fitting the red part of the spectrum with a combination of principal components (see \citealt{Paris2011}). We need to define the continuum in a wider wavelength range
($900 \mbox{\AA} < \lambda < 1200 \mbox{\AA}$ in the restframe of the quasar) compared to what was done by \citet{Paris2011}, they 
reconstructed continuum between the Ly-$\beta$ and C\,{\sc iv} emission lines.
To expand the continuum to the region $\lambda < 1025$ \AA\, we used a power law extrapolation of the mean 
quasar continuum (used in PCA method) adding features which account for the Ly-$\beta$ emission line and the Lyman break cutoff.
 
In the second method the spectrum is smoothed iteratively.
For each iteration we smoothed the spectrum by convolution with a Gaussian 
function with FWHM about 50~\AA. Then pixels deviating from the continuum by more than 3$\sigma$ are excluded at each iteration. The remaining pixels were used in the next iteration. We used four iterations, which was enough for convergence.

We found that the first method tends to overestimate the continuum and sometimes yields an incorrect shape in the presence of sharp features, 
while the second method tends to underestimate the continuum and to smooth out emission lines. Therefore, the final continuum was 
constructed as the geometric average of the two continua convolved with a Gaussian function 
with FWHM$\sim$1000 km/s (an example of reconstructed continuum is shown in Fig.~\ref{SDSSspectrum}). Obviously, the continuum reconstruction in the region bluewards the Ly-$\alpha$ emission line 
is rather difficult and sometimes ambiguous. Nevertheless, simulations presented in Section~\ref{test} show that the continuum reconstructed 
by the described procedure does not introduce any strong bias in the search for H$_2$ absorption systems.
For each spectrum in the samples we determined the signal to noise ratio (S/N) as the mean of S/N in each pixel 
over the wavelength range 1120$-$1040 \AA~ in the restframe of the DLA system. 

\subsection{The searching procedure}
\label{criteria}
\begin{figure*}
\begin{center}
        \includegraphics[width=1.0\textwidth]{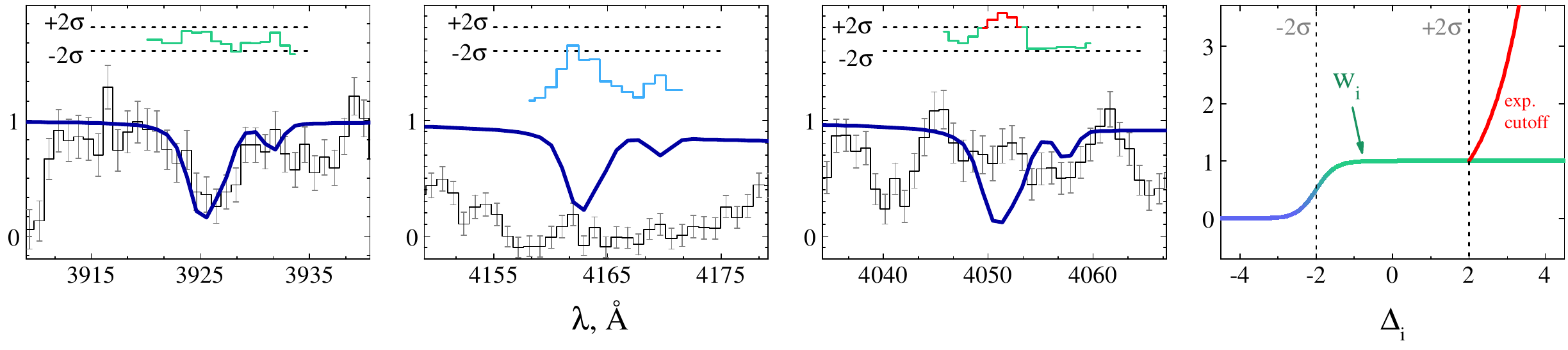}
        \caption{\rm Illustration of the searching criterion. The three left panels show different situations: 
respectively from left to right, \textit{good fit, blended and bad fit}. 
The blue line shows typical profiles of highly saturated (e.g. $\log$$N$$_{\rm H_2} = 20$) R(0), R(1) and P(1) lines convolved with 
the SDSS instrument function. The right panel shows the weight function and exponential cutoff function, that we used in our searching 
procedure (see equations \ref{Likely} and \ref{chi}). }
        \label{Criteria}
\end{center}
\end{figure*}

\noindent
Three main difficulties complicate the identification of H$_2$ absorption systems in SDSS spectra: 
(i) the intermediate spectral resolution, (ii) the usually low 
S/N and, (iii) the presence of the Ly-$\alpha$ forest. These difficulties jointly can lead to false identifications of H$_2$ absorption 
systems. The SDSS spectral resolution is the most important of them. The pixel size in the blue part of the SDSS spectrum is about 1~\AA. 
It sets a lower limit on the equivalent width of the H$_2$ lines which can be identified in SDSS spectra. The condition 
that H$_2$ equivalent widths to be larger than 1~\AA~ is fulfilled only for H$_2$ column density larger than $\log$$N$${\rm_{H_2}}$ $\gtrsim 18.5$.  The molecular hydrogen absorption spectrum in the UV appears as a series of absorption lines 
corresponding to Lyman and Werner vibrational bands L$\nu$-0 and W$\nu$-0. Intermediate SDSS spectral resolution leaves the 
possibility to observe only the most prominent lines corresponding to the R(0), R(1) and P(1) transitions in each band. These three absorption 
lines for each band are overlapped with each other under such resolution. 
We will refer to these absorption features by the name of the band, i.e. in case L4-0R(0), L4-0R(1) and L4-0P(1) are overlapped, 
the absorption feature will be denoted as L4-0 absorption line. 
The typical shapes of these lines are shown in Fig.~\ref{Criteria} by blue color.

We will construct a template of H$_2$ absorption systems to fit the SDSS spectra.
The H$_2$ absorption lines (from low rotational levels) have restframe wavelengths $\lambda<1115$ \AA. In principle, before the Lyman cutoff (at 912 \AA) 
of the DLA system, 26 absorption lines (from the Lyman and Werner bands) can be observed. However, H\,{\sc i} 
absorption lines of high order Lyman series of the DLA system can substantially absorb the spectrum blueward of $\sim$950 \AA. 
At the SDSS resolution, it leaves only about 17 H$_2$ lines which can be identified in the spectrum. 
However, the number accessible H$_2$ absorption lines varies for different spectra primarily because of the difference between the DLA and quasar redshifts and also because of the spectrum quality (in SDSS spectra S/N usually decreasing with decreasing wavelength). If the quasar redshift is low, the blue end of the spectrum limits the number of fitted H$_2$ lines. 
If the quasar redshift is high ($z>2.8$), then the presence of Lyman limit systems or saturated Ly$\alpha$ forest lines at redshift 
between $z_{\rm QSO}$ and $z_{\rm DLA}$ can reduce the number of H$_2$ lines in the fit.
To characterize the number of fitted lines we determined the cutoff wavelength $\lambda_{\rm cutoff}$ as the position in the 
spectrum from which the flux exceeds the zero level at the 2$\sigma$ level over more than 4 pixels in a row. In the following we will use 
the parameter $\lambda_{\rm B} = \lambda_{\rm cutoff}/(1+z_{\rm DLA})$ -- the cutoff in the spectrum expressed in the DLA restframe. 
This parameter just characterizes the number of H$_2$ absorption lines, that was used in the fit. The pixels corresponding to the DLA
Lyman series lines (Ly$\beta$,  Ly$\gamma$, etc) were excluded from the fit.

To identify H$_2$ absorption systems we used the profile fitting technique which compares a real spectrum with a H$_2$ model spectrum with specified physical parameters (i.e. column density, Doppler parameter, redshift). As we have to search H$_2$ systems in the large number of spectra ($\sim$ 13\,000), we need to develop an automatic procedure. 
We constructed a function based on the standard $\chi^2$ likelihood function commonly used in profile fitting procedures. 
This function is defined as
\begin{equation}
	\log L = \frac{\sum_i w_i\xi_i}{\sum_i w_i},
	\label{Likely}
\end{equation}
where we introduced the weight $w_i$ for each pixel
\begin{equation}
	w_i (\Delta_i) = \frac{1}{1+e^{-4(\Delta_i+2)}},
	\label{weight}
\end{equation}
where $\Delta_i = (y_i-f(x_i))/\sigma_i$ is the relative deviation of the model $f_i$ from the observed flux $y_i$ in pixel $x_i$, 
$\sigma_i$ is the error in pixel $x_i$. The weight function $w_i$ 
was chosen in the following way (see right panel of Fig.~\ref{Criteria}). We rejected those pixels (i.e. their $w_i = 0$) for which the spectrum is more than two $\sigma$ below the model
because they are possibly related to blends which are numerous in the Ly-$\alpha$ forest. For pixels where 
$2> \Delta_i > -2$, $w_i \approx 1$. 
The function $\xi_i$ is given by
\begin{equation}
\xi_i = \left\{
   \begin{array}{l l}
     \Delta_i^2e^{(1-\Delta_i)^2-1} & \quad \Delta_i \ge 2\\
     \Delta_i^2 & \quad \Delta_i < 2\\
   \end{array} \right\},
   \label{chi}
\end{equation}
where for $\Delta_i \ge 2$, an exponential cut-off is introduced because in case H$_2$ is present, all the absorption lines should be consistently seen.
Obviously, it is reasonable to fit only the absorbed part of the spectrum. Therefore the sum is taken only over the pixels where the model 
$f$ is less than 0.9 times the continuum level. 
Note, that the introduction of the exponential factors implies that $L$ defined by Eq.~\eqref{Likely} is not a likelihood function. 
The function $L$ is used only to select H$_2$ system candidates. We will estimate the confidence level of the detection in Sect.~\ref{confidence}.
We set the criterion of identification as $L < L_{id}$, where $L_{id}$ is the value of $L$ with $\Delta_i = 2$ for all pixels ($w_i \approx 1$ when $\Delta_i = 2$) at certain redshift. This criterion is fulfilled in case of the satisfactory fit of spectrum by H$_2$ absorption system model, i.e. there are no outliers in pixels, where observed flux is sufficiently larger than fit flux.

In standard profile fitting the best fit is obtained by minimization of a likelihood function over the parameter space. However, such a process requires 
to calculate a large number of absorption profiles which in our case would be particularly computational demanding.
Rather we constructed a set of H$_2$ absorption system template spectra on a dense grid of total column density and effective excitation 
temperature. These two parameters mainly define the line profiles of saturated H$_2$ absorptions at a given spectral resolution. The total 
H$_2$ column density ranges from log~$N$(H$_2$)~=~18.5 to 21 by step of 0.1 dex. With column densities less than 18.5 the absorption lines become 
weak and reliable identification of H$_2$ is impossible (see Sect.~\ref{ControlRate}).  
The effective excitation temperature, $T_{\rm 01}$, specifies the relative populations of H$_2$ J=0 and J=1 levels. $T_{\rm 01}$ was varied 
from 25 to 150 K corresponding to the typical observed range \citep{Srianand2005}.
The Doppler parameter is not important since we considered only highly saturated absorption systems ($\log{ N}> 18.5$). An important
ingredient is the velocity structure of the absorption system, 
i.e. the number of components, their relative strengths, and positions. High resolution observations show that more than 50 per cent 
of the H$_2$ absorption systems are multicomponent. However, the typical velocity separation of these components is less than 
50 km/s which is smaller than the SDSS spectral resolution ($\sim 150$ km/s). 
Therefore, we use a single component model. 
This implies however that our inferred H$_2$ column densities can be overestimated. This is partially confirmed by the analysis of the H$_2$ 
systems previously identified in high resolution spectra (see Sect.~\ref{candidates}).

We have implemented a fully automatic searching procedure. For each spectrum in the S$_{\rm DLA}$ sample  we calculated $L$ function in the redshift window for each model of H$_2$ absorption system in the template. 
The redshift window was taken as $\pm$600 km/s around the redshift of the DLA system. The DLA redshifts were taken from \citet{Noterdaeme2009, Noterdaeme2012}. Note that the value 600 km/s is larger than the typical observed velocity dispersion for H$_2$ absorption systems in DLAs. For each H$_2$ absorption system model from the template we have searched for redshifts at which identification criterion $L<L_{id}$ is satisfied.
Note that function $L$ guarantees that if H$_2$ absorption system with column density $N_0$ is satisfied $L<L_{id}$ for some position $z$ then 
H$_2$ absorption system with column density $N<N_0$ (less saturated) will also satisfied $L<L_{id}$ at $z$. 
 
Applying our searching procedure to the S$_{\rm DLA}$ sample we have selected a preliminary S$_{\rm cand}$ sample  of H$_2$ system 
candidates. For each candidate in the sample we have recorded the largest total column density for which $L < L_{id}$ is satisfied and the value 
of $z_{0}$ where $L$ has minimum for this largest total column density. The preliminary sample of candidates, S$_{\rm cand}$, contains over 4\,000 records. However most of these candidates are false detections caused mainly by Ly-$\alpha$ forest features in low S/N spectra. Indeed the observed occurrence rate of H$_2$ candidate in DLA based on the preliminary sample ($\sim$4000/13000=0.3) is larger than the $<$0.1 estimated rate based on high resolution data \citep{Noterdaeme2008}. To select most promising candidates for follow-up spectroscopic studies we need to select only candidates with high 
confidence, i.e. the candidates with a low probability of false detection. Two methods to estimate the probability of false detection are 
described in Sect.~\ref{confidence}.

\subsection{Testing the searching procedure}
\label{test}
\noindent

We need to be convinced that the overwhelming majority of real H$_2$ absorption systems will be detected by the searching procedure. 
One of necessary conditions is that the known H$_2$ absorption systems should be detected by our procedure in SDSS data.
There are only five such systems with SDSS data. They are listed in Table~\ref{table_known} and will be discussed in 
Section~\ref{candidates}. Only three of these spectra are useful
because of the proper spectral wavelength range. The searching procedure detects H$_2$ in all of them. This is fine but not enough to be convinced of the robustness of the searching procedure. 
Therefore we have tested the searching procedure on the data where we artificially added H$_2$ systems. For this we have used the S$_{\rm nonDLA}$ sample (where there are no DLA systems, see Sect.~\ref{data}) and artificially added DLA systems with one H$_2$ component with 
a specified column density $N$ at a given redshift. We used each spectrum several times varying the DLA redshift to increase statistics.
With this procedure we are sure that we have the same situation of noise and redshift distribution as in our search sample.

\begin{figure}
\begin{center}
        \includegraphics[width=0.47\textwidth]{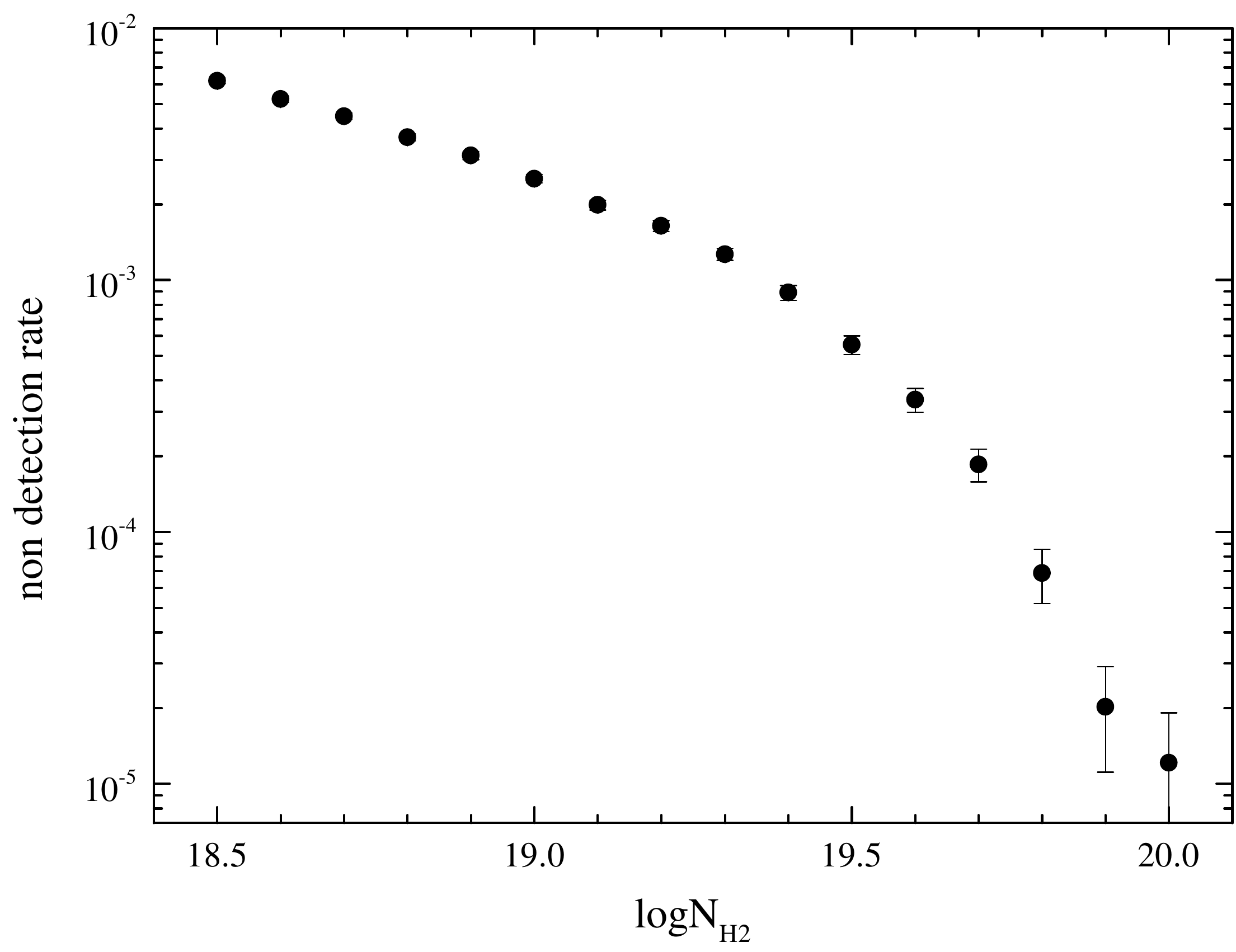}
        \caption{\rm Non-detection rate of the searching procedure versus log~$N$(H$_2$).
This rate is found to be less than 1\%.
}
        \label{nondetect}
\end{center}
\end{figure}

After applying our searching procedure we find that the typical non-detection rate (i.e. the fraction of systems we miss) is less than 1\%. The dependence of the non-detection rate with the specified column density is shown in Fig.~\ref{nondetect}. 

We conclude that our searching procedure gives a sample which is complete at the $>99\%$ level for log~$N$(H$_2$)~$>$~18. However, as we will show in the following section at such column densities the false identifications dominate the candidate sample and the limit column density for reliable identification is log~$N$(H$_2$)~$>$~19. For column density log~$N$(H$_2$)~$>$~19 our searching procedure gives completeness at more than $99\%$ level.

\section{Confidence estimation}
\label{confidence}

In the previous Section we have shown that our procedure is able to identify strong H$_2$ absorption systems 
in SDSS spectra when they are present. However, at this resolution, the numerous Ly-${\alpha}$ forest lines 
can easily mimic an H$_2$ absorption system and lead to false identifications. Therefore, 
in order to select a robust sample of H$_2$ system candidates,
we need to estimate the confidence level of the candidates, i.e. the probability for an H$_2$ candidate 
to be real. For this we determined a false identification probability (FIP).
It is the probability to wrongly select an H$_2$ absorption system in some realization of the Ly-$\alpha$ forest. If FIP is small, then the selected H$_2$ absorption system is likely to be real, i.e. its confidence value is high. 

We have applied two methods for FIP determination. 
In the first method for each possible detection, we have calculated the probability that a similar absorption system is detected anywhere else in the spectrum (see Sect.~\ref{MonteCarlo}).
In the second method we have calculated the rate of H$_2$ absorption system identification in the samples 
of spectra where there are confidently no H$_2$ absorption systems, i.e. in control samples. The latter method 
presented in Sect.~\ref{ControlRate} allows us in addition to estimate a detection threshold of H$_2$ absorption 
system in SDSS spectra and the detection probability of H$_2$ absorption systems in DLAs. 

\subsection{Monte-Carlo sampling}
\label{MonteCarlo}

The main point of this method is H$_2$ absorptions to be numerous, so as a typical system will be detected
by about 6-15 absorption features.
Suppose that the probability to ``fit'' by chance one H$_2$ line in the forest is 0.5. 
This means that the absorption line satisfies the identification criterion ($L<L_{id}$) over half of 
any wavelength window. Then the chance to ``fit'' $N$ lines together would be about $(0.5)^N$. 
The joint probability to fit a H$_2$ system by chance in a forest where no H$_2$ system is present
can thus be very small.
Therefore it is reasonable to expect that we can identify H$_2$ absorption systems even at the SDSS resolution 
and spectral quality.

To estimate the confidence of each candidate in the S$_{\rm cand}$ sample we used the following procedure. We performed 
arbitrary {\sl random} shifts of the H$_2$ absorption lines detected in the candidate. The shift of each line was limited by 
the position of the adjacent H$_2$ lines in the spectrum, typically less than 4000 km/s. In other words, we randomly 
``shake'' the H$_2$ absorption line positions. 
We performed many realizations and for each one we calculated the value of 
the $L$ criterion. The identification probability is estimated as:
$$
f_{\rm FP} = n(L<L_{id})/n_{\rm all},
$$
where $n(L<L_{id})$ is the number of identifications, i.e. realizations that satisfied ($L<L_{id}$), and $n_{\rm all}$ is the total number of 
realizations. The value of $f_{\rm FP}$ can be simply described as the probability of joint fit of several absorption lines in the
particular realization of the Ly-$\alpha$ forest in the spectrum. If $L<L_{id}$ for the majority of realizations, then $n(L<L_{id})$ is close to $n_{\rm all}$ and 
$f_{\rm FP}$ is close to 1. In this case Ly-$\alpha$ forest can effectively mimic the absorption system and the identification
cannot be considered as robust. If $L<L_{id}$ in a few of many realizations, then $f_{\rm FP}$ approaches 0. In this case the 
probability of the Ly-$\alpha$ forest to mimic the absorption system is small and the confidence in the identification of the system is high.
It should be emphasize that this method takes naturally into account the peculiarities of each spectrum.

\begin{figure*}
\begin{center}
        \includegraphics[width=1.0\textwidth]{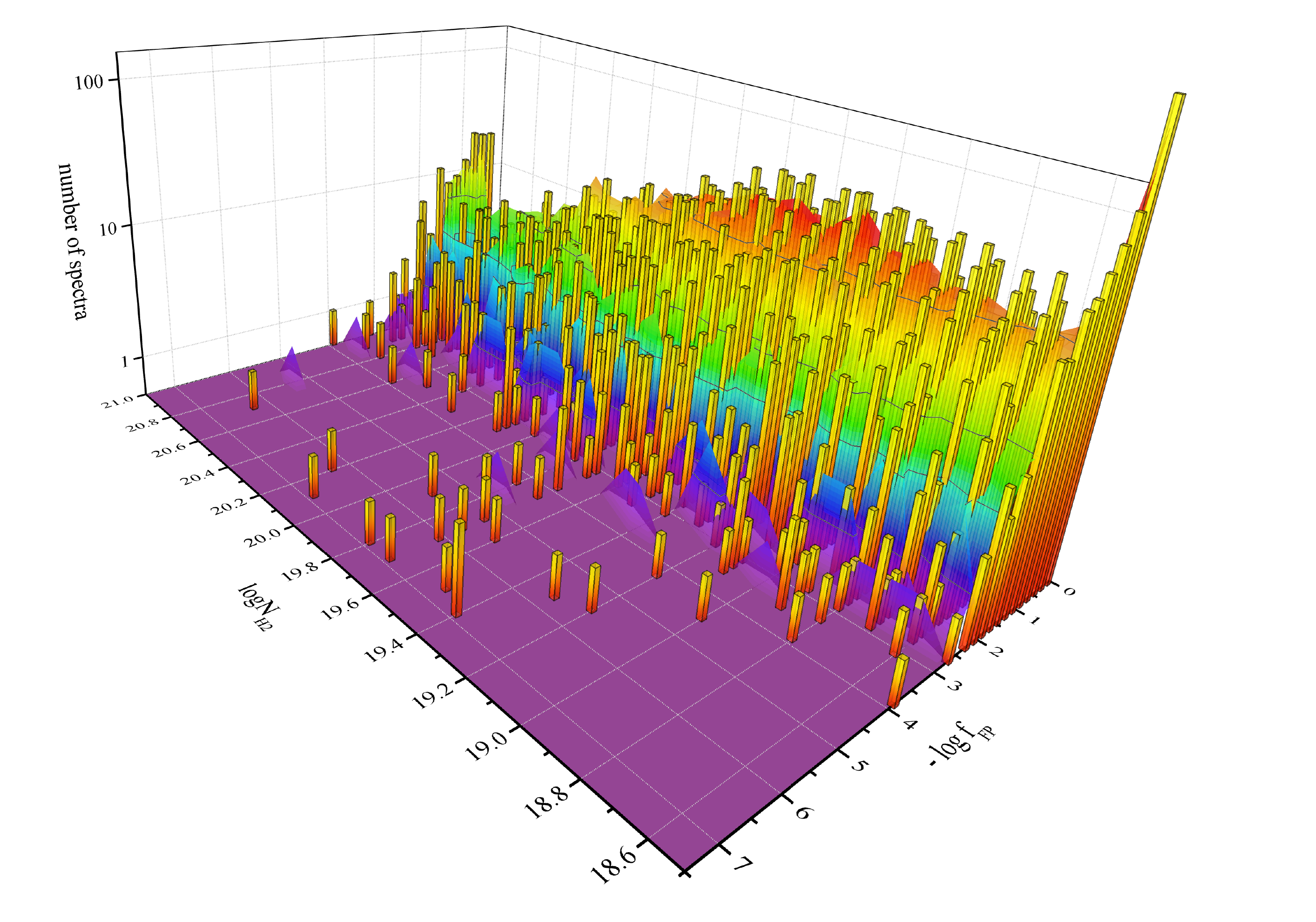}
        \caption{\rm The distributions of $f_{FP}$ values versus the column density of the H$_2$ absorption system candidates. 
The orange columns show the calculated distribution using the S$_{\rm DLA}$ sample (where H$_2$  systems are to be found), 
while smooth profiles show the distribution calculated from the S$_{\rm nonDLA}$ control sample (where no H$_2$ systems
are expected). The value of $\log(f_{\rm FP})=-3$ can be considered as the 
threshold for robust identification, since there are only a few spectra from the S$_{\rm nonDLA}$ control sample 
with smaller $f_{\rm FP}$ values.}
        \label{Limit}
\end{center}
\end{figure*}

The value of $f_{\rm FP}$ gives an estimate of the FIP for the candidate. 
However H$_2$ absorption lines have rigid relative position, while in the above calculation of $f_{\rm FP}$ we used random 
relative positions to increase statistics. We note that applying random shifts to H$_2$ absorption lines with respect to a fixed Ly-$\alpha$ forest is equivalent to applying random shifts to Ly-$\alpha$ forest lines with respect to fixed H$_2$ absorption. The latter would be difficult to realize technically, therefore we shifted H$_2$ absorption lines from their positions. Nevertheless it is necessary to estimate the $f_{\rm FP}$ value at which 
the detection can be considered as robust. To do this we used the following steps. We have applied the searching procedure 
to the SDSS spectra without DLA systems. We chose subsample, S$'_{\rm nonDLA}$, of the control sample S$_{\rm nonDLA}$ with size equal to the size of S$_{\rm DLA}$ sample. For each spectrum with z$_{\rm QSO}$ from the S$'_{\rm nonDLA}$ subsample we have generated the redshift of a ``fictitious'' DLA system using the 
distribution of z$_{\rm DLA}$ in the subsample of S$_{\rm DLA}$ spectra which have QSO redshifts in the range z$_{\rm QSO}\pm0.1$. The absence of real DLA systems in the spectra of S$'_{\rm nonDLA}$ subsample guarantees that 
there is no H$_2$ absorption system in the spectrum. 
We searched for ``H$_2$ candidates'' in this sample which are almost certainly false identifications. For each 
of the selected ``H$_2$ candidate'' we have calculated $f_{\rm FP}$. The distributions of $f_{\rm FP}$ values for 
the S$_{\rm DLA}$ (indicated as discrete measurements) 
and S$'_{\rm nonDLA}$ (smooth profiles) samples are shown in Fig.~\ref{Limit} versus the H$_2$ column densities. 
It is apparent that a value of $\log f_{\rm FP} < -3$ 
can be considered as the limit for reliable identification, as there is sufficient excess in the distribution of the candidates 
from the S$_{\rm DLA}$ sample.

\subsection{Use of a control sample}
\label{ControlRate}
The FIP of an H$_2$ absorption system can be estimated as the rate of identifications in spectra where no
H$_2$ absorption system is expected. 

FIP mainly depends on the quality of the spectrum, density of the Ly-$\alpha$ forest and certain profile of H$_2$ absorption system. We took four parameters (which we found enough) to describe this dependence: $z_{\rm QSO}$, S/N, $N_{\rm H_2}$ 
and the number of H$_2$ absorption lines that can be seen in the spectrum. We characterize the latter by the parameter 
$\lambda_{\rm B} = \lambda_{\rm cutoff}/(1+z_{\rm DLA})$ which is the blue limit of the spectrum expressed in the DLA restframe. 
H$_2$ absorption lines have rest-wavelength $\lambda<$1110\AA~. 
In order to detect at least one H$_2$ line, $\lambda_B$ must be $<$ 1110~\AA.
FIP has to be determined at each combination of these parameters. We have calculated FIP 
on a grid of H$_2$ column densities, S/N, $\lambda_{\rm B}$ and z$_{\rm QSO}$ parameters values. 
Log(S/N) was considered from 0.2 to 1.4 by step of 0.1. 
The different values of $\lambda_{\rm B}$ were taken to correspond to an increment in the 
number of H$_2$ lines. The redshift of QSO, $z_{\rm QSO}$, was varied in the range from 2.2 to 4.2 with 0.4 step. 

\begin{figure*}
\begin{center}
        \includegraphics[width=1.0\textwidth]{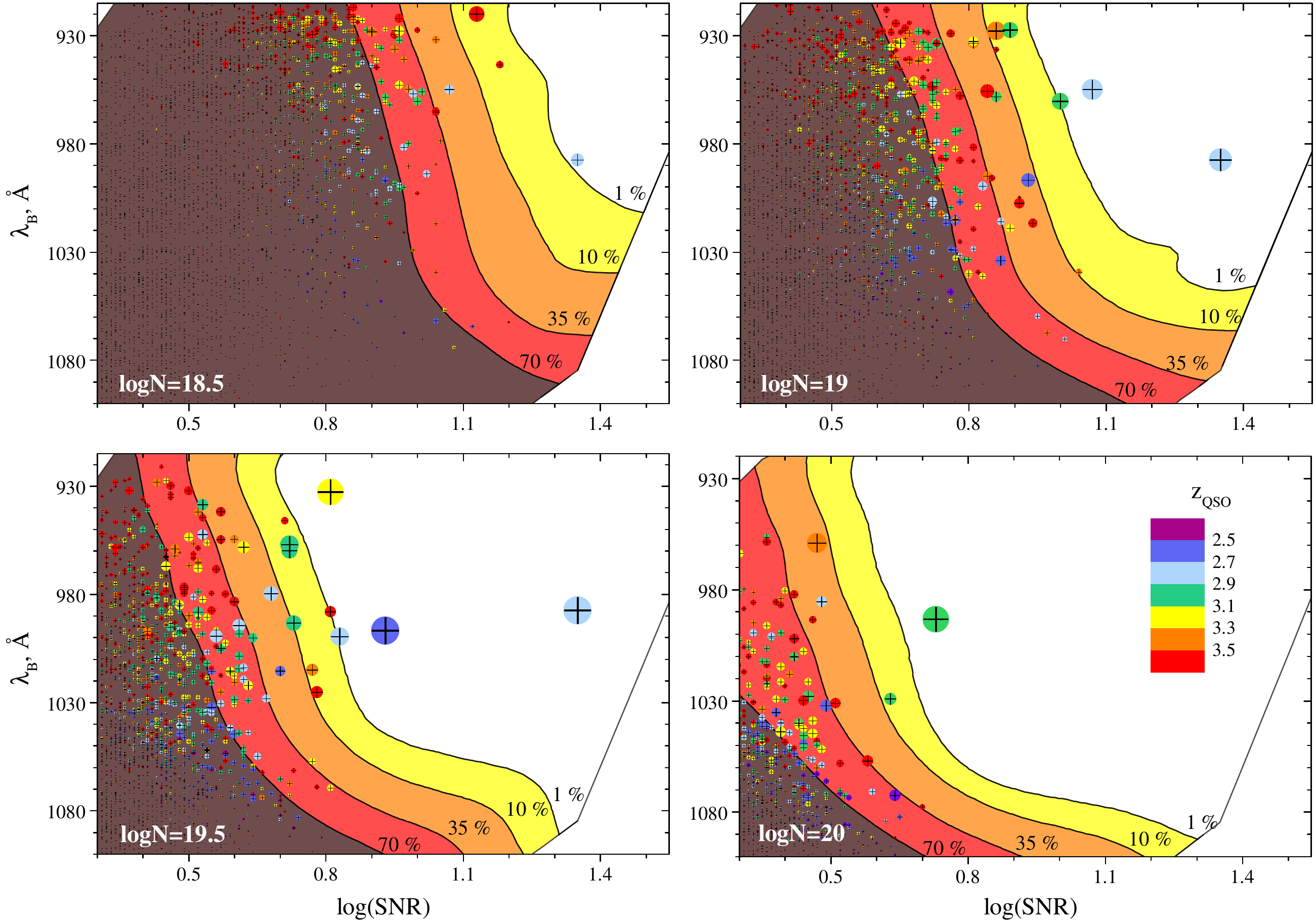}
        \caption{\rm Contour plots of false identification probability of H$_2$ absorption systems estimated by using the control 
sample S$'_{\rm nonDLA}$.  The different panels show $f_{\rm CS}$ for different $\log N$ - the total H$_2$ column density.
Y-axis correspond to $\lambda_{\rm B}$ - a parameter characterizing the number of H$_2$ bands available. 
X-axis corresponds to the signal-to-noise ratio of the spectrum. The white, yellow, orange, red and brown regions 
correspond to FIP values $<$1\%, $<$10\%, $<$35\% and $<$70\%, respectively. The black crosses show our
candidates H$_2$ absorption systems.
The size of each cross corresponds to the value of $f_{\rm FP}$ estimated from the Monte-Carlo method. 
The bigger size, the lower $f_{\rm FP}$ value. The color of each cross corresponds to the  redshift of the candidates.}
        \label{Probab}
\end{center}
\end{figure*}

\begin{figure}
\begin{center}
        \includegraphics[width=0.47\textwidth]{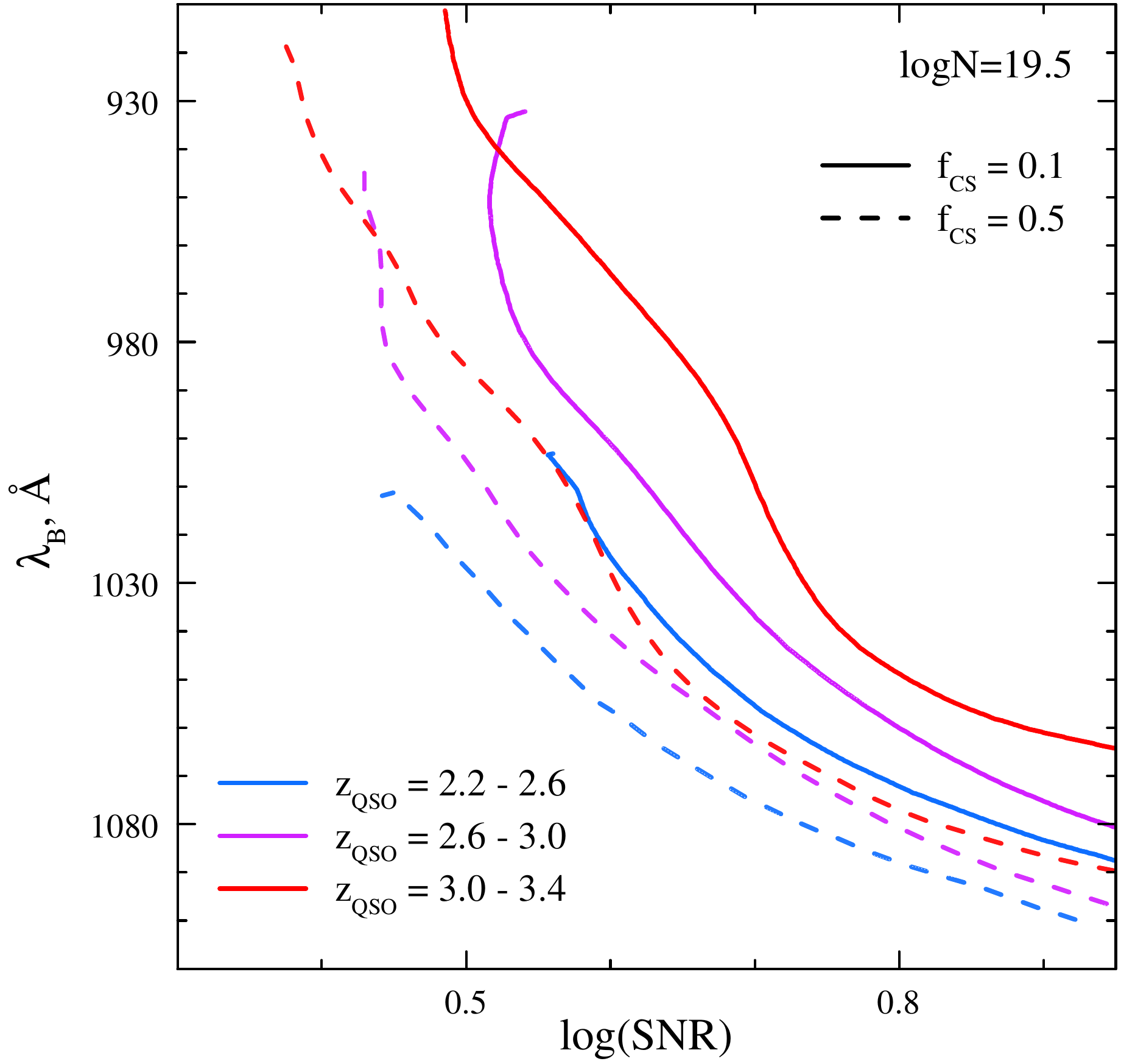}
        \caption{\rm Dependence of the $f_{\rm CS}$ (the upper limit on the false identification probability) on $z_{\rm QSO}$. 
Solid and dashed lines show the $f_{\rm CS}$ isocontours at $0.1$ and $0.5$, respectively. The $f_{\rm CS}$ is calculated for 
log~$N$(H$_2$)(cm$^{-2}$)~=~19.5. Blue, purple and red lines correspond to $f_{\rm CS}$ calculated using spectra with 
$z_{\rm QSO}$ in the ranges ($2.2\div2.6$),  ($2.6\div3.0$) and ($3.0\div3.4$), respectively.}
        \label{zdepend}
\end{center}
\end{figure}

For each spectrum in the S$_{\rm nonDLA}$ control sample 
we chose randomly several positions of $z_{\rm DLA}$. 
Each of z$_{\rm DLA}$ for the spectrum corresponds to different number of H$_2$ absorption lines involved in analysis, or $\lambda_{\rm B}$ bin. Therefore one spectrum can be used several times with different z$_{\rm DLA}$. In other words, we constructed enlarged sample S$'_{\rm nonDLA}$, which allowed us to sufficiently increase statistics. Then we have applied the searching procedure (see Sect~\ref{criteria}) to enlarged sample S$'_{\rm nonDLA}$ and have calculated the identification rate for each H$_2$ column density in the selected S/N and $\lambda_{\rm B}$ bins
\begin{equation}
f_{\rm CS}(\lambda_{\rm B}, S/N, {\rm z_{QSO}}, N_{\rm H_2}) = n(L < L_{id})/n_{\rm bin},
\end{equation}
$n(L < L_{id})$ is the number of spectra in the bin, for which the identification criterion ($L < L_{id}$) is satisfied, $n_{\rm bin}$ is the total
number of spectra in the bin. The identification rate in the S$_{\rm nonDLA}$ sample gives us the estimate of the FIP. The contour plots of $f_{\rm CS}$ for different $N_{\rm H_2}$ are shown on Fig.~\ref{Probab}. 
$f_{\rm CS}$  depends mainly on  $\lambda_{\rm B}$ and S/N and little on $z_{\rm QSO}$.
Therefore to construct the contour plots we integrated $f_{\rm CS}$ over the $z_{\rm QSO}$ bins. 
The dependence of FIP, $f_{\rm CS}$, on $z_{\rm QSO}$ is  shown in Fig.~\ref{zdepend}. 
Note that the contour plots shown on Figs.~\ref{Probab} and \ref{zdepend} are smoothed over the bins. The general behavior of 
$f_{\rm CS}$ satisfies reasonable expectations. The probability decreases with decreasing $\lambda_B$ (i.e. increase of the number 
of available H$_2$ lines), and with increasing S/N and decreasing $z_{\rm QSO}$ (forest less dense). 
Additionally, the probability decreases with increasing total H$_2$ column density N$_{\rm H_2}$. 

The calculated identification rate $f_{\rm CS}$ gives an upper limit on the FIP of H$_2$ absorption system. Indeed, we searched H$_2$ system in some redshift window around z$_{\rm DLA}$. The probability that at some redshift in the search window identification criterion to be satisfied increase with increase of search window width. Therefore the rate $f_{\rm CS}$ is scaled with changing the search window width.  The rate $f_{\rm CS}$ approaches FIP value with reducing search window width. However, we can't set search window very small since we do not know the exact position of H$_2$ system relative to z$_{\rm DLA}$ (which sets the center of the search window).

To estimate the false identification probability of our H$_2$ candidates derived from the S$_{\rm cand}$ sample we have to position the corresponding spectra on Fig.~\ref{Probab}. The candidates can be seen as crosses.
The size of each cross indicates the calculated false identification probability from the Monte-Carlo method (see sect.~\ref{MonteCarlo}). Note that the FIP estimated from the two methods are in agreement.
It is apparent that the overwhelming majority of candidates are located in regions where the FIP probability is high, i.e. these identifications 
are not robust. However, some spectra are located in regions with low false identification probability. 

Using these considerations, we can estimate the thresholds in S/N and $\lambda_B$ for robust identification at a given $N_{\rm H_2}$
at some $f_{\rm CS}$ level.
We note that the detection limit of H$_2$ absorption systems in SDSS spectra is higher than $\log$N$\sim 19$. Indeed, it is seen from Fig.~\ref{Probab} that for confident detection, $f_{\rm CS} < 10\%$, of H$_2$ absorption system with column density $\log$N$\sim 19$ the high signal to noise spectrum is required. Such spectra are very rare in SDSS database.

Sample S$'_{\rm nonDLA}$ has a limited number of spectra in some bins of the parameter space, especially at high S/N and low $\lambda_B$ value. 
Additionally we have used each spectrum several times. Therefore, we used simulated spectra 
to check the influence of these limiting factors.
We generated a S$_{\rm Ly\alpha}$ sample quasar spectra.
The overall shape of the continuum was constructed using principal component coefficients from \citet{Paris2011} 
(we used only the 10 first PCA coefficients).
The Ly$\alpha$ forest column density and Doppler parameter distributions and the evolution with redshift of the number density were taken from \citet{Meiksin2009}. 
The initial spectrum was calculated at high resolution ($>$ 100,000) and then convolved with the SDSS BOSS instrumental function 
taken from \citet{Smee2013}. 
Finally, we added Poisson noise corresponding to the SDSS sample.
We generated a substantial number of spectra ($\sim$ 400) in each bin of $\lambda_B$ and S/N. We applied our searching 
procedure and calculated the identification rate in each bin. We find that $f_{\rm CS}$ estimated from the SDSS DR9 sample and 
from the mock sample generally are agree within statistical errors. 

\subsection{Probability of H$_2$ detection in DLAs}

\begin{figure}
\begin{center}
        \includegraphics[width=0.47\textwidth]{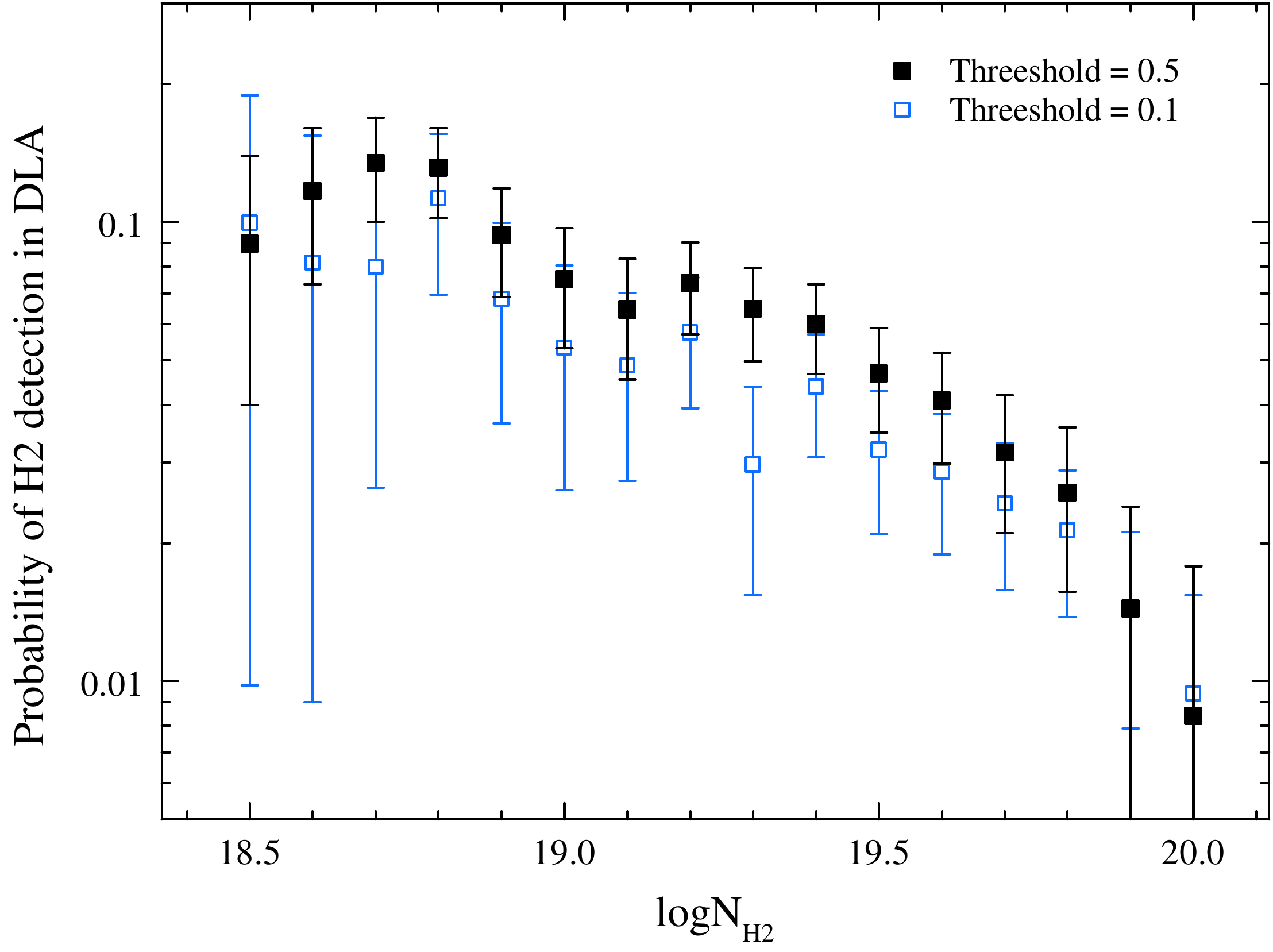}
        \caption{Detection probability of H$_2$ absorption systems in DLAs versus column density. 
Note that this probability should be considered as an upper limit because our H$_2$ column densities are generally 
overestimated. Blue open and black filled squares correspond to the calculation with 0.1 and 0.5 threshold values of the detection rate, respectively.}
        \label{probability}
\end{center}
\end{figure}

The FIP estimated in the previous section allows us to determine the probability of H$_2$ detection in DLA systems. For each column density 
we have selected the regions of the $\lambda_B$-S/N parameter space (shown on Fig.~\ref{Probab}) where $f_{\rm CS}$ is less than a
given threshold value. In principle it is better to use as low threshold value as possible, because this reduces possible biases
and increases the robustness of the detection. However, in order to increase statistics, we have taken a threshold value of $50\%$. 
We checked that lower threshold values give consistent results, with larger statistical errors.
In each bin of the $\lambda_B$-S/N parameter space we compared the identification rates of H$_2$ absorption systems in 
the S$_{\rm DLA}$ and  S$'_{\rm nonDLA}$ samples. The probability of detection of H$_2$ 
systems with column density larger than a given value $N_{\rm H_2}$ can be calculated as
\begin{equation}
	P({\rm N_{H_2}})	= \frac{\sum{N_{S_{\rm DLA}} \times (f_{\rm CS}(S_{\rm DLA}) - f_{\rm CS}(S'_{\rm nonDLA}))}}{\sum{ N_{S_{\rm DLA}}}},
\end{equation}
where sums are taken over the bins with $f_{\rm CS}(S'_{\rm nonDLA})$ less than some threshold value, and $N_{S_{\rm DLA}}$ is the number of spectra with DLA systems in 
the bin, $f_{\rm CS}(S_{\rm DLA})$ and $f_{\rm CS}(S'_{\rm nonDLA})$ are the detection rates of H$_2$ absorption systems in the bin in the $S_{\rm DLA}$ 
and $S'_{\rm nonDLA}$ samples, respectively. The results of the calculation for two threshold values are shown in Fig.~\ref{probability}. The higher threshold values give larger statistics and therefore smaller error bars. However, calculation using higher threshold values lead to higher values for obtained probability, which is seen in Fig.~\ref{probability}. Nevertheless, the results of the calculation for threshold values 0.1 and 0.5 are in the agreement with each other.
Note that this probability, especially at the low column densities, should be considered as an upper limit because our procedure tends to
overestimate the column density due to the quality of SDSS spectra. 

Based on the VLT survey of H$_2$ absorption systems in
QSO spectra \citep{Noterdaeme2008} it was found that
the overall covering factor of H$_2$ in DLAs is $\sim$10\% for $\log f>-4.5$.
For the high end of the N(H$_2$) as considered here, the VLT survey
indicates rather 8\% (N(H$_2$)$>$18) or less, in very good agreement
with our analysis.

\section{Candidates}
\label{candidates}
\noindent

\begin{figure*}
\begin{center}
        \includegraphics[width=1.0\textwidth]{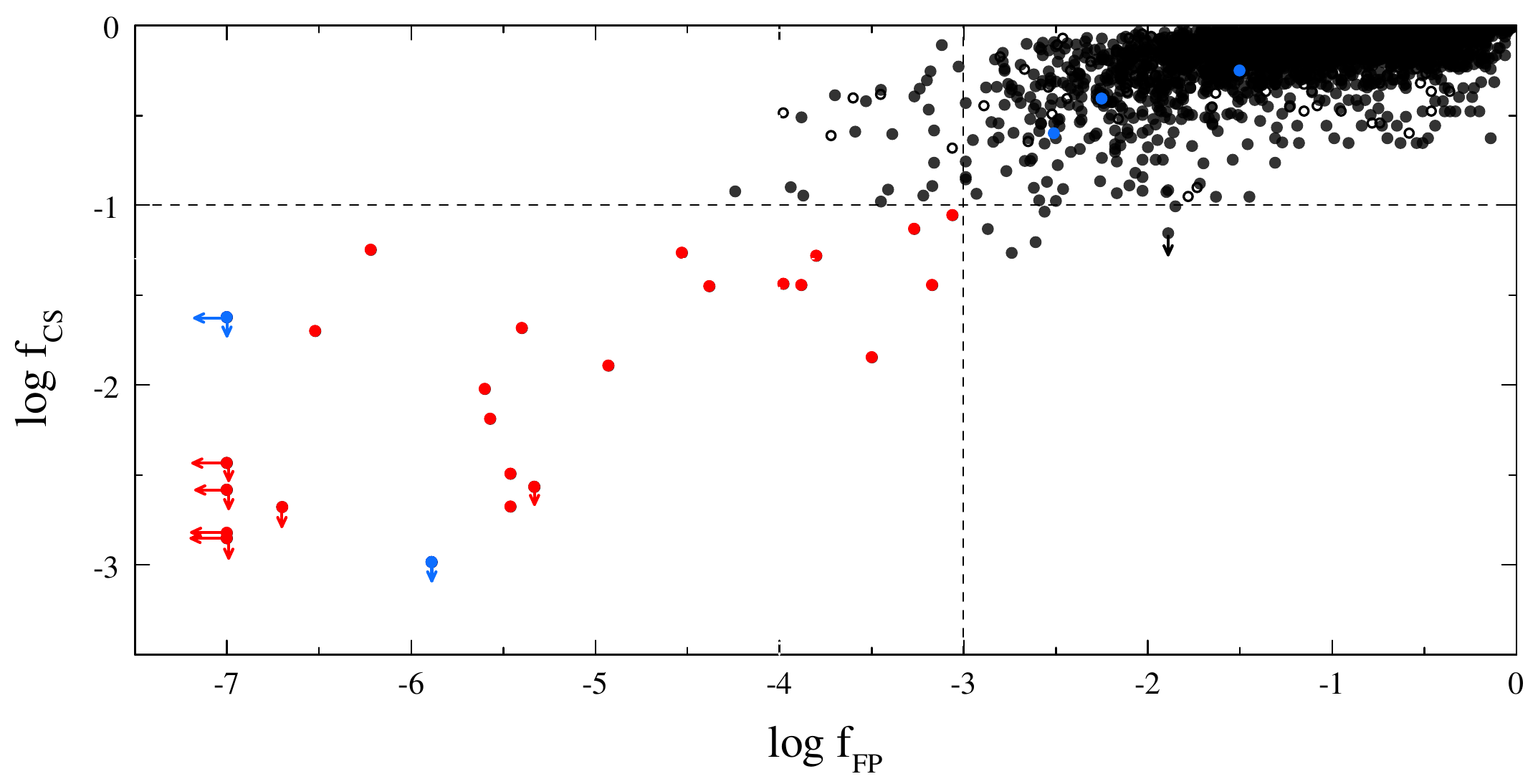}
        \caption{\rm The outputs of the two methods used to estimate the confidence level of H$_2$ candidates are used
to select the final sample. 
Red circles show selected H$_2$ candidates. Filled and open circles correspond to candidates from DR9 and DR7, respectively.
Blue circles show the known H$_2$ absorption systems. Two of them (J081240.69+320808.52 and J123714.61+064759.64) are easily found by our method.} 
        \label{comparison}
\end{center}
\end{figure*}

We have compiled the final sample, S$_{\rm final}$, of the most promising H$_2$ candidates for follow up studies.  Fig.~\ref{comparison} shows 
the $f_{\rm FP}$ and $f_{\rm CS}$ values calculated for each candidate in the S$_{\rm cand}$ sample by the two methods presented in 
Sect.~\ref{MonteCarlo} and ~\ref{ControlRate}, respectively. As we stated above $f_{\rm CS}$ calculated using the control sample gives an upper limit on the FIP, as it depends on the width of the search window. In turn $f_{\rm FP}$ estimated using Monte-Carlo simulations gives a lower limit on FIP. Nevertheless it is seen again that Monte-Carlo sampling and control sample methods are in 
agreement with each other. The S$_{\rm final}$ sample presented in Table~\ref{table_cand} is formed from new H$_2$ candidates which have 
$\log f_{\rm FP}<-3$ (see Fig.~\ref{Limit}) and $f_{\rm CS}<0.1$. The selected candidates are shown in Fig.~\ref{comparison} by red circles. The known H$_2$ absorption systems shown in Fig.~\ref{comparison} by blue points were excluded from the S$_{\rm final}$ sample and not listed in Table~\ref{table_cand}.
The filled and open circles in Fig.~\ref{comparison} correspond to candidates from DR9 and DR7, respectively. 
We found that improved quality of quasar spectra in SDSS-III has major importance for detection of H$_2$ absorption systems.   
There are only two candidates (J081240.69+320808.52, J153134.59$+$280954.36) in DR7 satisfied the selection criterion for S$_{\rm final}$ sample 
($\log f_{\rm FP}<-3$ and $f_{\rm CS}<0.1$). They were both reobserved and presented DR9 catalog thus we used its DR9 spectra. 

\begin{table*}
         		\begin{tabular}{|l|c|c|c|c|c|c|c|c|c|c|c|}
            \hline
            Quasar \quad & Plate-MJD-Fiber & $z_{em}$ & $z_{DLA}$ & $\log$N$_{\rm H\,I}$ & S/N & $\lambda_B$, \AA  & $\log$N$_{\rm H_2}$ & T$_{01}$  & $log(f_{\rm FP})$ & $f_{\rm CS}$ & $\Delta_V$\\
            \hline
J234730.76$-$005131.68 &4214-55451-0212 & 2.63 & 2.5874 & 20.20 & 0.9 &  996 & 19.5 & 25 & $<$-7.0 &      0.1 & -60  \\
J221122.52$+$133451.24 &5041-55749-0374 & 3.07 & 2.8376 & 21.85 & 0.7 &  993 & 20.1 & 25 & $<$-7.0 & $<$  0.1 & -70  \\
J114824.26$+$392526.40 &4654-55659-0094 & 2.98 & 2.8320 & 20.84 & 1.0 &  960 & 19.4 & 25 & $<$-7.0 & $<$  0.3 & -30  \\
J075901.28$+$284703.48 &4453-55535-0850 & 2.85 & 2.8221 & 20.87 & 1.1 &  955 & 19.2 & 25 & $<$-7.0 & $<$  0.4 & -70  \\
J153134.59$+$280954.36 &3959-55679-0862 & 3.23 & 3.0025 & 21.01 & 0.8 &  932 & 19.5 & 25 & -6.7 & $<$  0.2 & 60  \\
J001930.55$-$013708.40 &4366-55536-0874 & 2.53 & 2.5284 & 20.64 & 0.5 & 1032 & 20.7 & 25 & -6.5 &      2.0 & -130  \\
J152104.92$+$012003.12 &4011-55635-0218 & 3.31 & 3.1410 & 21.56 & 0.5 &  958 & 20.2 & 25 & -6.2 &      5.6 & -150  \\
J013644.02$+$044039.00 &4274-55508-0691 & 2.81 & 2.7787 & 20.47 & 0.7 &  979 & 19.7 & 25 & -5.6 &      1.0 & -110  \\
J105934.34$+$363000.00 &4626-55647-0381 & 3.77 & 3.6402 & 20.97 & 0.8 &  955 & 19.3 & 25 & -5.6 &      0.7 & -80  \\
J164805.16$+$224200.00 &4182-55446-0880 & 3.08 & 2.9900 & 21.52 & 0.7 &  957 & 19.6 & 25 & -5.5 &      0.3 & -40  \\
J160638.54$+$333432.89 &4965-55721-0091 & 3.09 & 3.0845 & 20.42 & 0.9 &  927 & 19.2 & 25 & -5.5 &      0.2 & -80$^*$  \\
J123602.11$+$001024.60 &3848-55647-0266 & 3.03 & 3.0289 & 20.58 & 0.5 &  938 & 19.9 & 25 & -5.4 &      2.1 & -420$^1$  \\
J144132.27$-$014429.40 &4026-55325-0848 & 2.98 & 2.8908 & 21.46 & 0.7 &  959 & 19.7 & 25 & -5.3 & $<$  0.3 & 110$^*$  \\
J150739.67$-$010911.16 &4017-55329-0647 & 3.10 & 2.9743 & 20.05 & 0.6 &  958 & 19.8 & 25 & -4.9 &      1.3 & -30  \\
J150227.22$+$303452.68 &3875-55364-0040 & 3.33 & 3.2828 & 20.83 & 0.9 &  927 & 19.0 & 25 & -4.5 &      5.5 & 0  \\
J082102.66$+$361849.68 &3760-55268-0364 & 2.81 & 2.8030 & 20.38 & 0.6 &  994 & 19.7 & 25 & -4.4 &      3.5 & -10  \\
J084312.72$+$022117.28 &3810-55531-0727 & 2.91 & 2.7866 & 21.80 & 0.4 & 1040 & 21.0 & 25 & -4.0 &      3.6 & -10  \\
J123052.64$+$020834.80 &4752-55653-0116 & 3.36 & 2.7981 & 21.26 & 0.8 & 1015 & 19.7 & 25 & -3.9 &      3.6 & 20  \\
J082716.26$+$395742.48 &3761-55272-0810 & 2.83 & 2.7420 & 20.59 & 1.0 &  956 & 18.8 & 25 & -3.8 &      5.2 & 50  \\
J004349.39$-$025401.80 &4370-55534-0422 & 2.96 & 2.4721 & 20.57 & 0.6 & 1029 & 20.1 & 25 & -3.5 &      1.4 & 270$^2$  \\
J120847.64$+$004321.72 &3845-55323-0604 & 2.72 & 2.6083 & 20.35 & 1.0 &  993 & 18.8 & 25 & -3.3 &      7.4 & -40  \\
J160332.00$+$081622.44 &4893-55709-0876 & 2.86 & 2.8429 & 20.27 & 0.9 &  942 & 18.9 & 25 & -3.2 &      3.6 & -10  \\
J141205.80$-$010152.68 &4035-55383-0704 & 3.75 & 3.2678 & 20.53 & 0.9 & 1007 & 19.1 & 25 & -3.1 &      8.8 & 30  \\
  \hline
            \end{tabular}

            \caption{The S$_{\rm final}$ sample of new H$_2$ absorption system candidates in SDSS DR9. Here $z_{em}$, $z_{DLA}$ and $\log$N$_{\rm H\,I}$~(cm$^{-2}$) are quasar and DLA system redshifts and column density of H\,{\sc i} taken from \citet{Noterdaeme2009}, S/N is $\log$ of estimated signal to noise ratio of the spectrum in the region where
H$_2$ lines should be seen, $\lambda_B$ is wavelength of the cutoff in the spectrum in the DLA restframe, $\log$N$_{\rm H_2}$~(cm$^{-2}$) is estimated upper limit on the H$_2$ total column density, $T_{01}, {\rm (K)}$ is kinetic temperature which gives the relative population of J=0, J=1 H$_2$ rotational levels, $f_{\rm FP}$ is the probability of false detection estimated by Monte-Carlo sampling, see Sect.~\ref{MonteCarlo},  $f_{\rm CS}$ is the probability of false detection estimated by control sample expressed in per cent, see Sect.~\ref{ControlRate}, $\Delta_V$, km/s is the difference between redshifts of H$_2$ bearing component and DLA system (measured by metal lines).
$^*$ C\,{\sc i} metal lines are tentatively detected in these spectra. It is believed that C\,{\sc i} is a good tracer of H$_2$.
$^1$ in this DLA system H$_2$ bearing component associated with the second, less prominent component in metal line profile. 
$^2$ low ionization metal lines is not detected in this system. z$_{\rm DLA}$ estimated using C\,{\sc iv} and  Si\,{\sc iv} metal transitions. Visual inspection showed that this candidate is unlikely.
}
            \label{table_cand}
\end{table*}

Note, that the H$_2$ column densities given in Table~\ref{table_cand} are generally overestimated. This arises mainly from two factors.
The first is the numerous blends in the Ly$\alpha$ forest which can be unresolved at the resolution of SDSS spectra. The second is that 
our procedure gives the highest column density for which the criterion of detection, $L<L_{id}$, is satisfied. 
We found that SDSS data quality is not high enough to estimate with reasonable uncertainty H$_2$ column densities using standard 
$\chi^2$ likelihood profile fitting. Using simulations of SDSS mock spectra with specified H$_2$ absorption systems we can estimate 
that the standard $\chi^2$ procedure gives systematic errors larger than 0.5~dex.

High resolution spectrum studies of H$_2$ absorption systems ($\log$~N~$\gtrsim19$) indicate that
H$_2$ bearing DLA systems usually are associated with prominent metal lines. 
We found that the redshifts of H$_2$ bearing component of the candidates is satisfied with redshifts of DLA system within 100 km/s, which is less than errors in the redshift determinations. Only in two of the H$_2$ candidates (see Table~\ref{table_cand}, they marked by an asterisk) we have detected C\,{\sc i} absorption lines which is known as a good tracer of molecules \citep{Srianand2005, Noterdaeme2012}. Detections of C\,{\sc i} are rare in SDSS spectra as a consequence of the low resolution of the spectra.

\begin{figure*}
\begin{center}
        \includegraphics[width=1.0\textwidth]{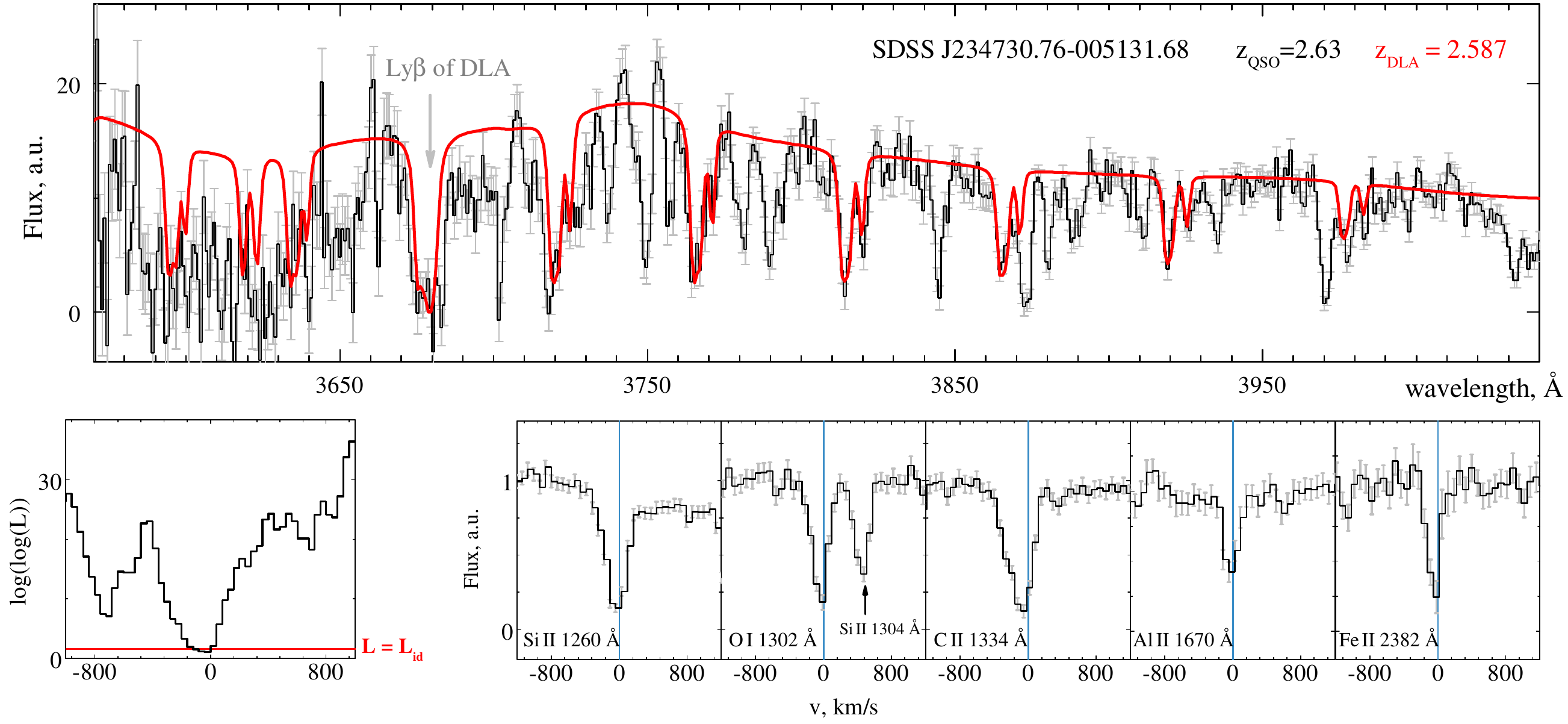}
        \caption{\rm The spectrum of SDSS J234730.76-005131.68. The top panel shows part of the spectrum with fitted H$_2$ profiles 
at z=2.587 overplotted. The five right bottom panels show metal lines associated with this system. 
Left bottom panel shows the dependence of the searching criterion with redshift, expressed as a velocity offset from the position of the system. }
        \label{J2347}
\end{center}
\end{figure*}

\begin{figure*}
\begin{center}
        \includegraphics[width=1.0\textwidth]{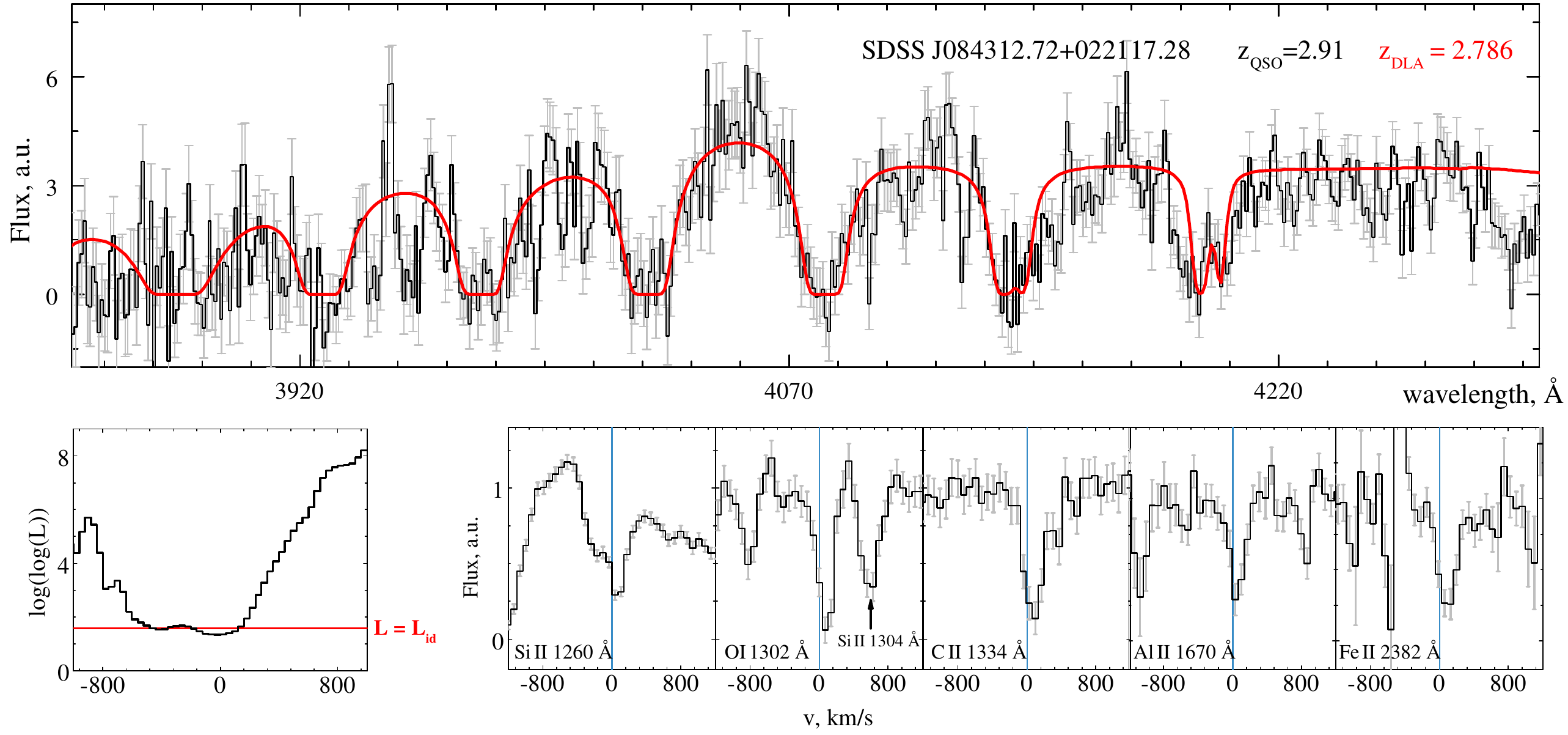}
        \caption{\rm The candidate in the spectrum of SDSS J084312.72+022117.28 
has possibly the highest column density ever detected in front of QSOs, $\log N_{\rm H_2} = 21.1$. The top panel shows part 
of the spectrum with fitted H$_2$ profiles. The five right bottom panels show metal lines associated with this system. 
Left bottom panel shows the dependence of the searching criterion with redshift, expressed as a velocity offset from the position of the system. }
        \label{J0843}
\end{center}
\end{figure*}

For instance, we present two candidates in the spectra of J234730.76-005131.68 and J084312.72+022117.28 
 in Fig.~\ref{J2347} and Fig.~\ref{J0843}, respectively. The  J234730.76-005131.68 spectrum has relatively high S/N and the candidate has very low $f_{\rm FP}$ value, i.e. the detection of the H$_2$ absorption system is robust. The J084312.72+022117.28 spectrum
has lower S/N and the candidate higher $f_{\rm FP}$ value, but from Fig.~\ref{J0843} it can be seen that highly saturated H$_2$ absorptions 
are prominent. The estimated column density of this H$_2$ candidate is $10^{21.1}$ cm$^{-2}$. 
Such high H$_2$ column densities have never been observed towards high redshift quasars. There are also 3 candidates in Table~\ref{table_cand}, which have estimated column densities exceeding $10^{20}$ cm$^{-2}$.
Such saturated H$_2$ systems are detected up to now only towards two GRBs \citep{Prochaska2009, Kruhler2013}, and are usual in our Galaxy \citep{Rachford2002}. These four selected highly saturated H$_2$ systems are very promising candidates to probe the translucent phase of high redshift ISM.

Table~\ref{table_known} gives the properties of seven H$_2$ absorption systems that were previously identified from high resolution spectra obtained at Keck and/or VLT observatories. One of them (J143912.04+111740.5) was not included in the DLA catalog due to poor S/N, but we included the spectrum in the searching 
sample, S$_{\rm DLA}$. For two of them H$_2$ lines fall out the searched region due to poor spectral quality. We identify the
H$_2$ absorption systems in all of the five remaining spectra. However the identification is robust only for two systems. They have $\log$N$_{\rm HR}$(H$_2$)~$>$~19 measured from high resolution data.
For the other three systems (J144331.17+272436.73, J081634.39+144612.36, 
J235057.87-005209.84) with H$_2$ column density $\log$N$_{\rm HR}$(H$_2$) lower than 19, our false identification probability is not low enough $>10^{-3}$, i.e. identification is not robust and they can not be considered as reliable candidates. 

\begin{table*}
         		\begin{tabular}{|c|c|c|c|c|c|c|c|c|c|}
            \hline
            Quasar \quad & {$z_{em}$} & {$z_{abs}$} & $\log$(S/N) & $log(f_{\rm FP})$ & $\log$~N~  & $\log$~N$_{\rm HR}$ & Instrument & comment & ref.\\
            \hline
J081240.69+320808.52 & 2.70 & 2.625 &  1.4 & $<$-7.0 & 19.8 &  19.88$\pm$0.06 & KECK/HIRES & HD & a \\
J123714.61+064759.64 & 2.79 & 2.689 &  0.8 & -5.5 & 19.9 &  19.21$^{+0.13}_{-0.12}$ &  VLT/UVES & multicomp, HD/CO & b\\
J144331.17+272436.73 & 4.44 & 4.225 &  0.6 & -2.5 & 19.2 & 18.29$\pm$0.08 & VLT/UVES &  & c \\
J081634.39+144612.36 & 3.85 & 3.287 &  0.6 & -1.7 & 19.8 & 18.66$^{+0.17}_{-0.30}$ & VLT/UVES &  multicomp & d\\
J235057.87-005209.84 & 3.02 & 2.425 &  0.6 & -1.4 & 19.7 & 18.52$^{+0.30}_{-0.49}$ & VLT/UVES & multicomp  & e\\
J143912.04+111740.5  & 2.58 & 2.418 &  0.6   & \multicolumn{2}{c}{Out of range}   & 19.38$\pm0.10$   & VLT/UVES    &   multicomp, HD/CO & f  \\
J091826.16+163609.0  & 3.07 & 2.58  &  0.1   & \multicolumn{2}{c}{Out of range}   &  19$\div$16   & VLT/X-shooter    &  not in DLA sample & g \\
            \hline
            \end{tabular}
            \caption{Already known H$_2$ system in SDSS. $log(f_{\rm FP})$ is a estimate of FIP, estimated by Monte-Carlo sampling method. $\log$~N is the H$_2$ column density estimated by our procedure. $\log$~N$_{\rm HR}$ is H$_2$ column density known from analysis of high resolution spectrum. Ref.: $^a$ \citet{Jorgenson2009}, $^b$ \citet{Noterdaeme2010}, $^c$ \citet{Ledoux2006}, $^d$ \citet{Guimaraes2012}, $^e$ \citet{Petitjean2006}, $^f$ \citet{Srianand2008}, $^g$ \citet{Fynbo2011}.}
            \label{table_known}
\end{table*}

\section{Properties of H$_2$ candidate sample}
\label{properties}
\noindent

\begin{figure}
\begin{center}
        \includegraphics[width=0.45\textwidth]{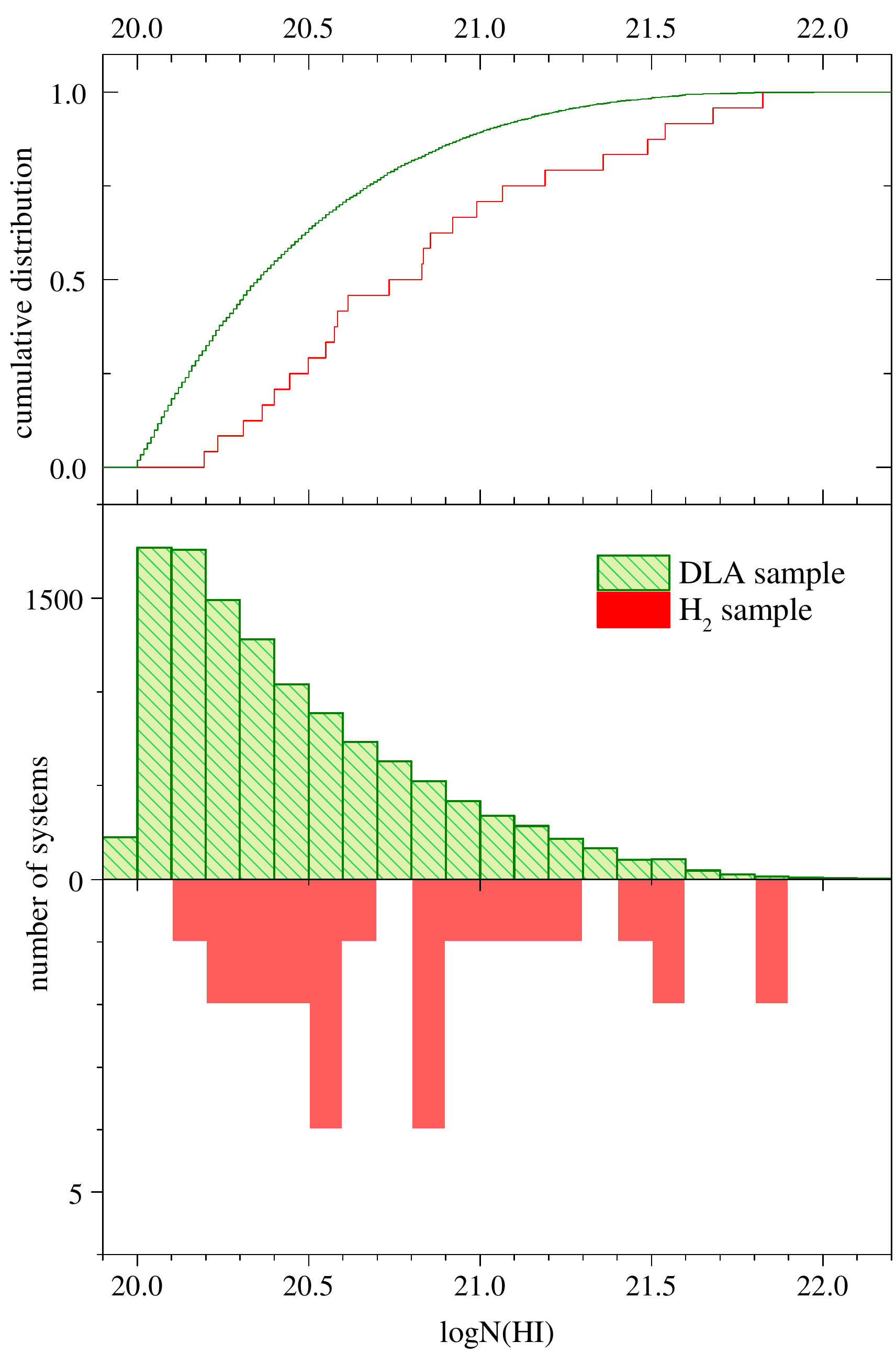}
        \caption{Distribution (bottom panel) and cumulative distribution (top panel) of the column densities of neutral hydrogen in the DLA systems. Red and green colors correspond to the sample of H$_2$ bearing candidates and S$_{\rm DLA}$ sample, respectively.}
        \label{HNIhist}
\end{center}
\end{figure}

In this Section we present the properties of candidates in the S$_{\rm final}$ sample. 
The main goal is to search for difference between the properties of H$_2$-bearing and non H$_2$-bearing DLA systems. The distributions of neutral hydrogen column density for H$_2$ candidate sample and DLA sample are shown in Fig.~\ref{HNIhist}. The Kolmogorov-Smirnov test indicates that we can reject the hypothesis that the two distributions are the same at the 0.99 level. Comparison of the distributions shows that the presence of high column density H$_2$ absorption systems lead to the higher H\,{\sc i} column densities.

\subsection{Colour excess}
\begin{figure*}
\begin{center}
        \includegraphics[width=1.0\textwidth]{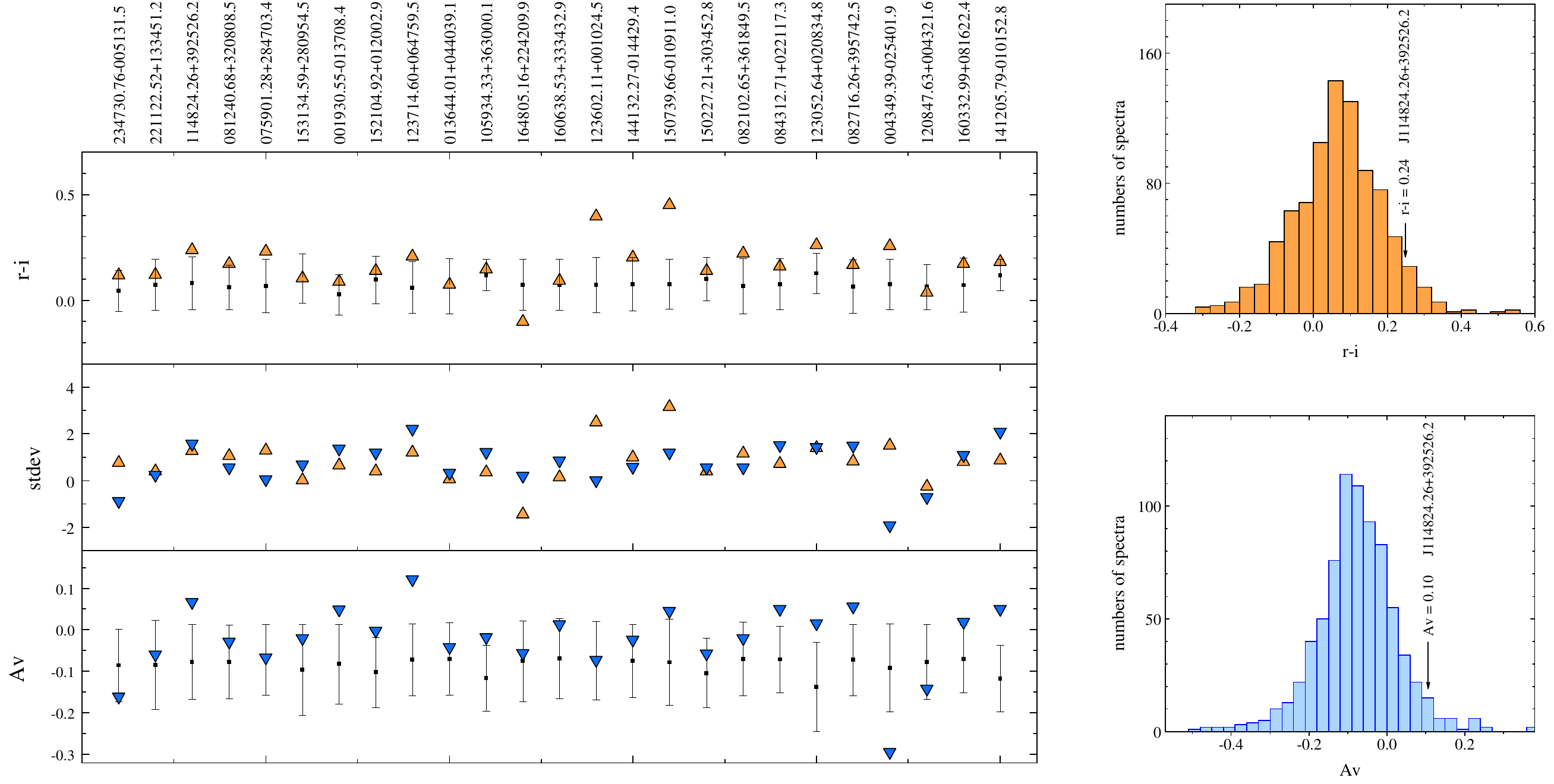}
        \caption{Estimated $r-i$ colors and $A_V$ for selected H$_2$ candidates. {\sl Top left}: the orange points in the top left panel show the $r-i$ color for H$_2$ candidates while the black points with error bars show the median value and dispersion of $r-i$ colors in control samples for each of the H$_2$ candidates. {\sl Bottom left}: the blue points in the top left panel show the $A_V$ for H$_2$ candidates while the black points with error bars show the median value and dispersion of $A_V$ in control samples for each of the H$_2$ candidates. {\sl Middle left}: orange and blue points indicate standard deviation of $r-i$ and $A_V$, respectively, measured for H$_2$ candidates from their control sample. {\sl Top right}: distribution of $r-i$ colour in the control sample of H$_2$ candidate in J114824.26+392526.2. {\sl Bottom right}: distribution of $A_V$ in the control sample of H$_2$ candidate in J114824.26+392526.2.}
        \label{g-r}
\end{center}
\end{figure*}

\begin{figure}
\begin{center}
        \includegraphics[width=0.47\textwidth]{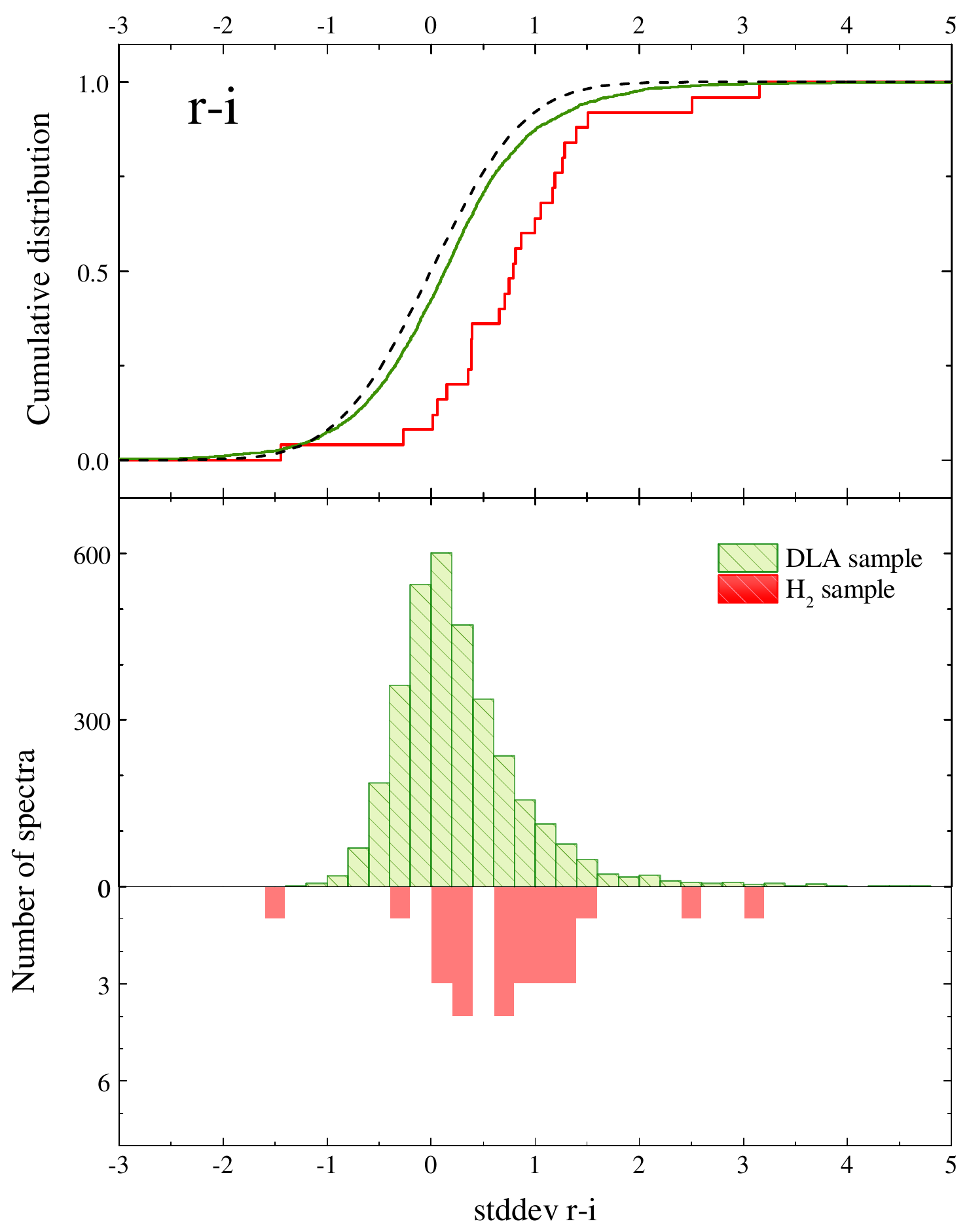}
        \caption{Distribution (bottom panel) and cumulative distribution (top panel) of the standard deviations of $r-i$ compared to the median in the control samples. Red and green colors correspond to the sample of H$_2$ bearing candidates and DLA sample, respectively.
The black dashed line in the top panel shows the cumulative probability for a normal distribution.}
        \label{colorhist}
\end{center}
\end{figure}

\begin{figure}
\begin{center}
        \includegraphics[width=0.47\textwidth]{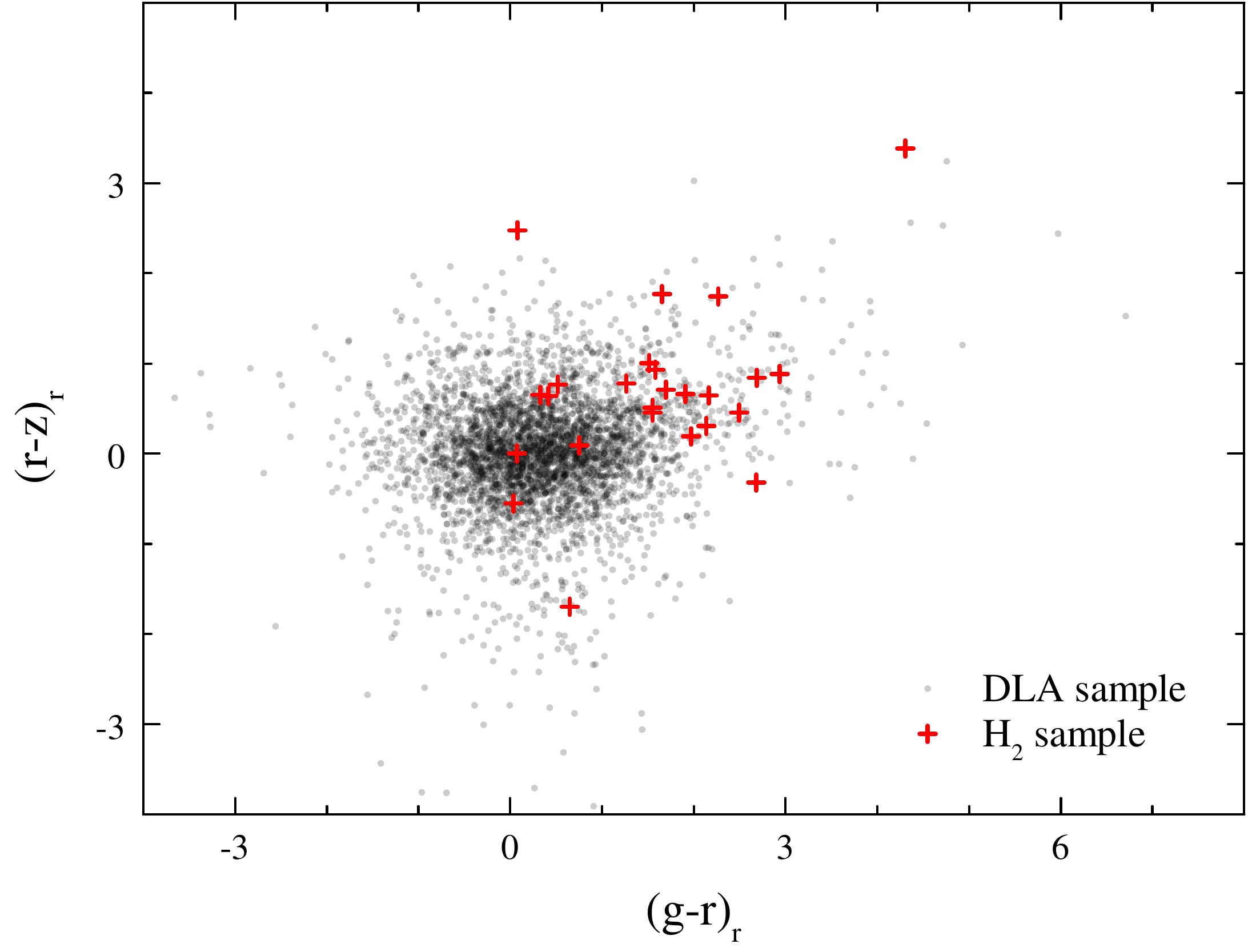}
        \caption{$(r-z)_r$ -- $(g-r)_r$ color diagram. Red and black points correspond spectra from the sample of H$_2$ bearing candidates and DLA sample, respectively. $(r-z)_r$ and $(g-r)_r$ values show $r-z$ and $g-r$ colors measured in standard deviations of their control samples.}
        \label{colorcolor}
\end{center}
\end{figure}

We have studied the colour excess for the H$_2$ candidate sample, S$_{\rm final}$. Recently it was shown that quasar spectra with 
H$_2$/CO-bearing absorption systems show significant $g-r$ colour excess compared to quasar spectra with non-molecular bearing
DLAs  (e.g. \citealt{Noterdaeme2010}). This excess might be caused by:
(i) the presence of H$_2$ absorption lines in H$_2$ candidate spectra;
(ii) the increased dust content of the H$_2$ bearing DLA (because H$_2$ molecules in the cold neutral medium mainly form onto the dust grains). To estimate the colour 
excess we used the following procedure. SDSS filters magnitudes were taken from the DR9 quasar catalog 
\citep{Paris2012}. For each H$_2$ candidate we have selected a control sample of quasars from the DR9 catalog with redshifts similar to redshift of the candidate
quasar (within $\Delta z =0.05$). We have compared the value of $r-i$ color
for each candidate with the median value of $r-i$ in the control samples (the top right and left panels of Fig.~\ref{g-r}). In comparison with previous studies we have chosen $r$ and $i$ filters because there is no bias due to the presence of H$_2$ absorption lines. Using dispersions measured in the control samples we have calculated standard deviations of the candidates from their control samples (the middle right panel of Fig.~\ref{g-r}). We have found 1$\sigma$ excess in the $r-i$ colors for H$_2$ candidate sample from their control samples. Additionally, we have found that $r-i$ color excesses for H$_2$ candidates are statistically higher than color excesses measured for the DLA sample (for this purpose we used the statistical DLA sample from \citet{Noterdaeme2012}, which is a subsample of S$_{\rm DLA}$ sample), see Fig.~\ref{colorhist}. Since H$_2$ absorption lines for H$_2$ candidate do not fall in $r$ and $i$ SDSS filters we suppose that this excess is the evidence of enhanced dust content in H$_2$-bearing DLAs. We found significant excesses of H$_2$ candidate sample over DLA sample in $g-r$, $u-r$, $r-z$ colours as well. The excesses in $g-r$, $u-r$ colors are about 2$\sigma$ of standard deviation, that are higher than $\sim 1\sigma$ excesses measured in $r-i$, $r-z$ (see Fig.~\ref{colorcolor}). It agrees with previous statement about higher extinction in H$_2$ bearing candidates, but can be also explained by location of H$_2$ absorption lines over $g$ and $u$ SDSS filters. 
The Fig.~\ref{colorcolor} shows comparison of standard deviations from their control samples of H$_2$ candidate and DLA samples, calculated for $g-r$ and $r-z$ colors. We suppose that measured colour excesses most likely are explained by the higher amount of dust in the H$_2$-bearing DLAs.
The latter can be investigated further  by extinction measurement from spectral fitting.

\subsection{Extinction}
\begin{figure}
\begin{center}
        \includegraphics[width=0.47\textwidth]{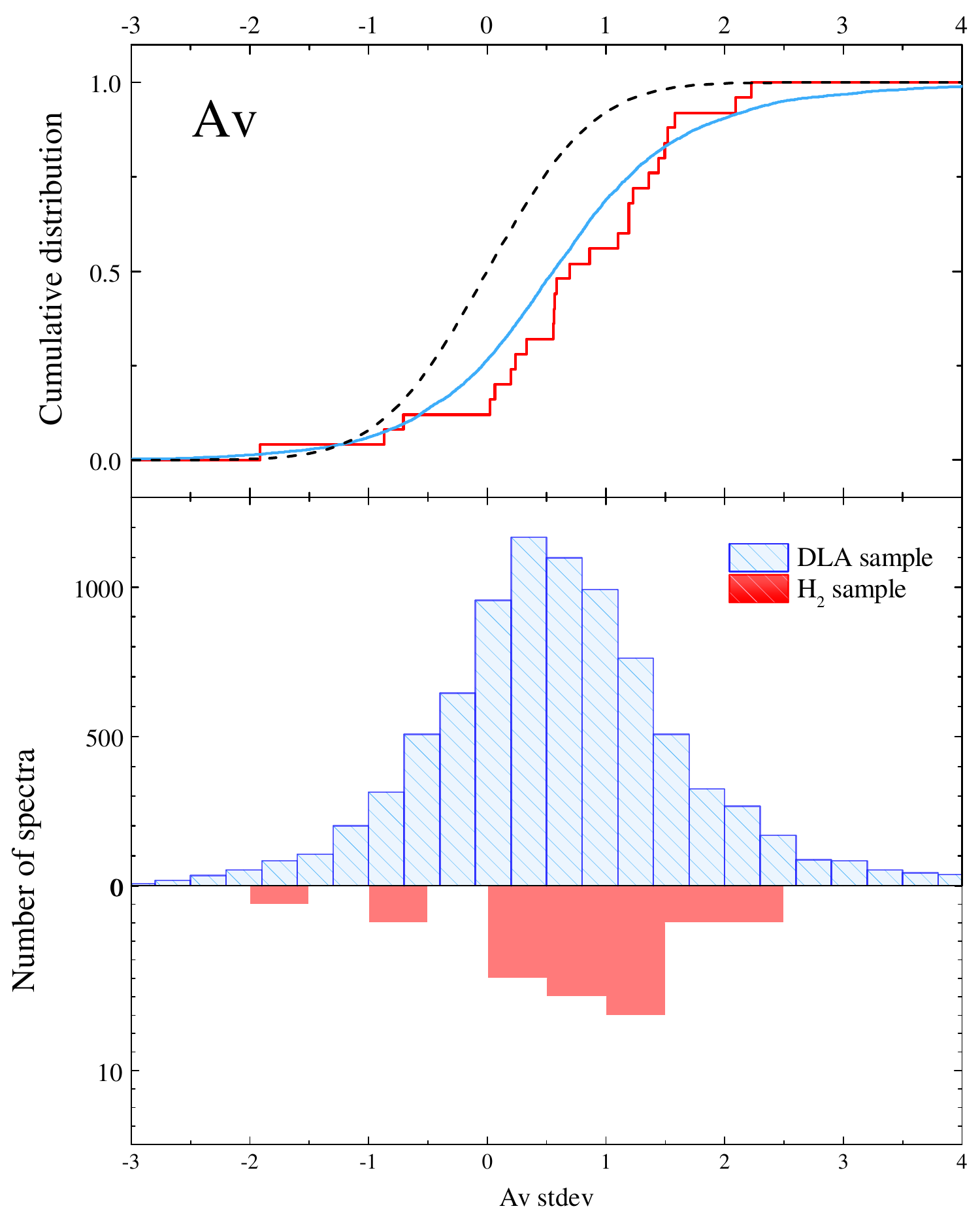}
        \caption{Distributions (bottom panel) and cumulative distributions (top panel) of the deviation (in unit of standard deviation) of $A_V$ 
compared to the mediam in the control samples.
Red and blue colors correspond to the sample of H$_2$ bearing candidates and the whole DLA sample, respectively. The black dashed line in 
the top panel show the cumulative probability for a normal distribution. A similar excess is seen for the two samples.}
        \label{Av_stdev}
\end{center}
\end{figure}

The 1$\sigma$ $r-i$ colour excess of H$_2$ candidates can indicate an enhanced dust content in H$_2$-bearing DLA systems.  
We therefore perform a direct measurement of extinction of the quasar spectra where H$_2$ candidates are found. We have used a procedure similar 
to that used by \citet{Srianand2008, Noterdaeme2009}. We have corrected the spectra for the Milky Way reddening using $A_V$ maps given in
\citet{Schlegel1998} and the improved correction formula by \citet{Schlafly2011}. We have fitted the QSO continuum of each candidate in the regions 
without emission lines. The positions of emission lines and initial QSO spectrum were taken from \citet{VandenBerk2001}. We used the SMC-like extinction 
curve which was applied at the DLA restframe. The normalization of the spectrum and $A_V$ were obtained during standard minimization $\chi^2$ fit. 
For each candidate we have constructed a control sample
including non-BAL quasars with no DLA and spectra of S/N$>3$ and redshifts within $\Delta z=0.1$ of the redshift of the 
H$_2$ bearing candidate QSO. For each quasar of the control sample we measure the virtual extinction that would produced by a DLA
at the redshift of the H$_2$ bearing candidate.
Each control sample has a $A_{\rm V}$ distribution close to normal (an example is shown in the right bottom panel of Fig.~\ref{g-r}). 
The measured median and dispersion of the $A_V$ distribution of the control sample for each candidate is shown as a black point with error bars 
in the left bottom panel of Fig.~\ref{g-r}. The measured $A_V$ values of the H$_2$ candidates are shown by blue triangles.
The distribution of the deviations (in unit of standard deviation) of $A_V$  in H$_2$ candidate spectra 
relative to the median in their control sample is shown in red color in Fig.~\ref{Av_stdev}. The same distribution calculated for the whole DLA 
sample, S$_{\rm DLA}$, is shown in blue color in Fig.~\ref{Av_stdev}. Although there is an excess of 
red color in the H$_2$  bearing sample, the Kolmogorov-Smirnov test indicates that we can not reject that the two distributions 
are the same (at the 0.41 level).  However, note that both samples have 1$\sigma$ from the 
control non-DLA non-BAL sample. If this excess is attributed to the dust presence in DLA systems, it implies median A$_V\lesssim 0.10$ for DLA systems as well as for H$_2$ bearing system. Such small values of extinction can not lead to any selection bias for DLA systems, which is in agreement with what was found in the radio selected QSO DLA surveys \citep{Ellison2001, Jorgenson2006}. 

\subsection{Metal content}
\begin{figure*}
\begin{center}
        \includegraphics[width=1.0\textwidth]{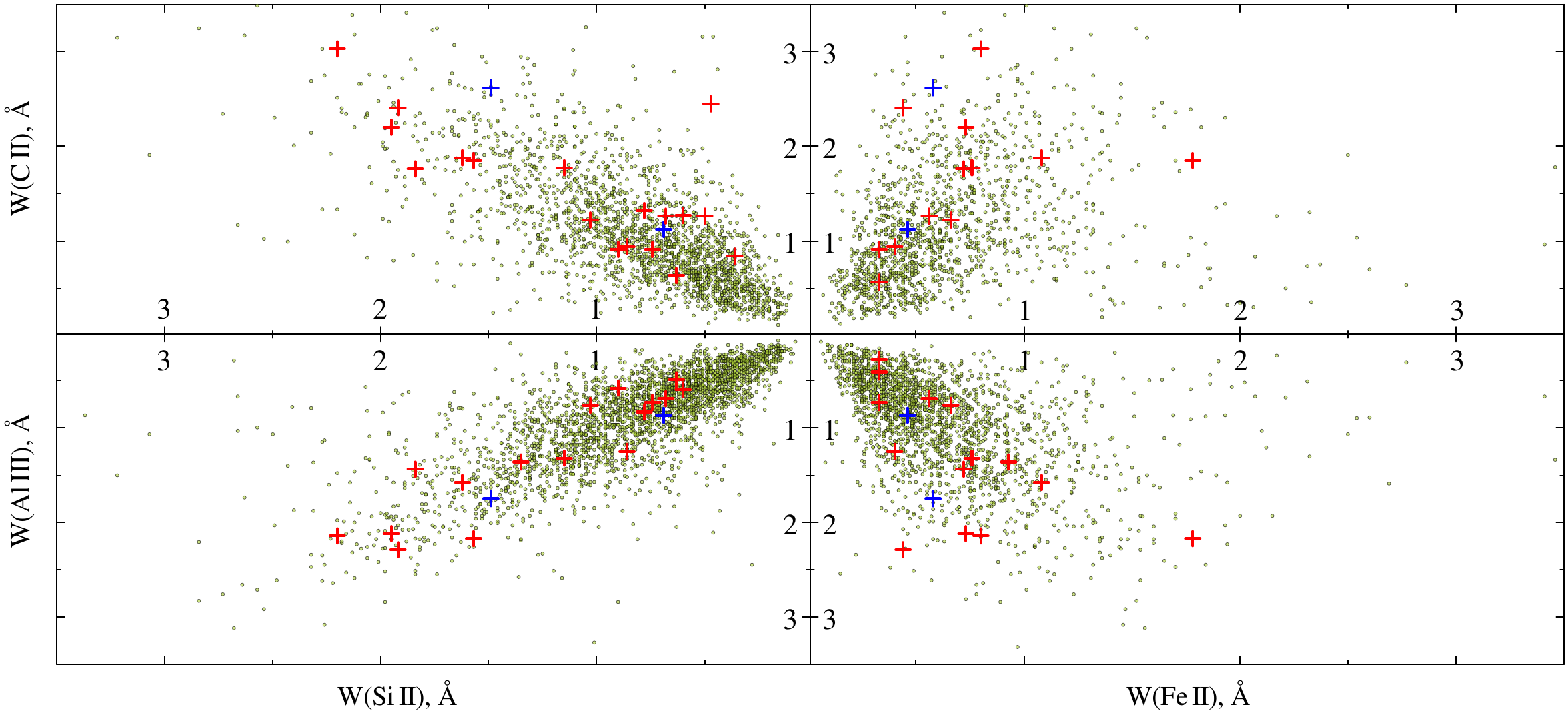}
        \caption{ Equivalent widths of  C\,{\sc ii}, Si\,{\sc ii}, Fe\,{\sc ii} and Al\,{\sc iii} absorptions. The green circles, red and blue crosses 
correspond to EWs for DLA systems detected in DR9, H$_2$ candidates and H$_2$ candidates confirmed by high resolution studies, respectively.}
        \label{metals}
\end{center}
\end{figure*}

\begin{figure}
\begin{center}
        \includegraphics[width=0.47\textwidth]{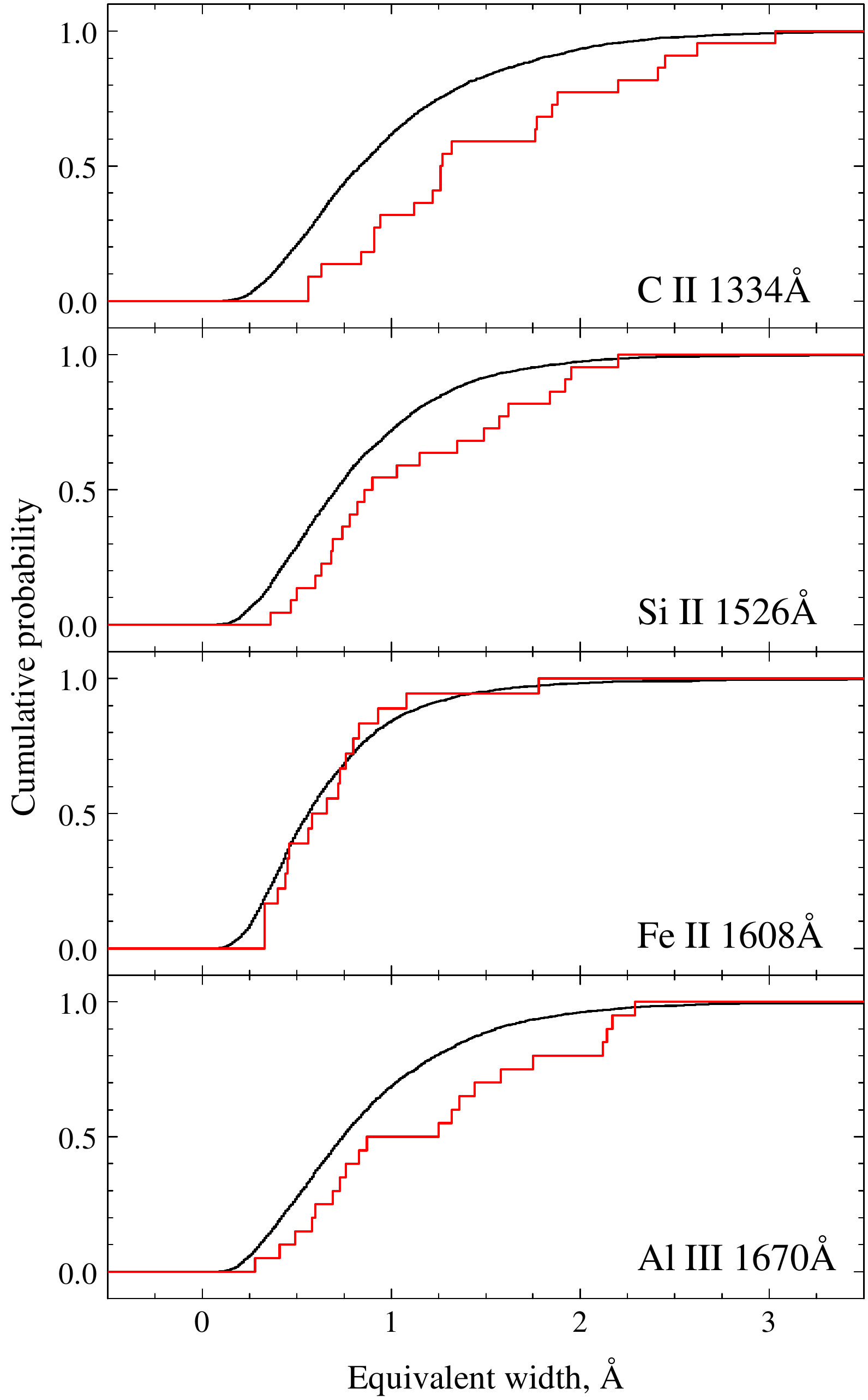}
        \caption{ Cumulative distribution of equivalent widths of C\,{\sc ii}, Si\,{\sc ii}, Fe\,{\sc ii} and Al\,{\sc iii} absorptions. The black and red lines in each panel 
correspond for samples of DLA systems detected in DR9 and H$_2$ candidates, respectively.}
        \label{metals_c}
\end{center}
\end{figure}

The measured equivalent widths (EW) of metal transitions can be used to characterize the
metal content in DLA systems. We used EWs of C\,{\sc ii} 1334\,\AA, Si\,{\sc ii} 1526\,\AA , 
Fe\,{\sc ii} 1608\,\AA, and Al\,{\sc iii} 1670\,\AA\,\,automatically measured in the DLA DR9 catalog \citep{Noterdaeme2012}. The distributions of 
EWs for DLA and H$_2$ candidate samples are 
shown in Fig.~\ref{metals} and Fig.~\ref{metals_c}. Note, to construct cumulative distributions in Fig.~\ref{metals_c} we used only spectra where metal lines are detected. It rejects most of possible false-positive DLA systems which presence in DLA DR9 sample. However, there can be a fraction of DLAs with low metal content, where some metal lines can't be detected in SDSS spectra. Using Kolmogorov-Smirnov test we have found that EWs in the H$_2$ candidate sample is higher than EWs in DLA sample for C\,{\sc ii} 1334\,\AA, Si\,{\sc ii} 1526\,\AA\,\,and Al\,{\sc iii} 1670\,\AA\,\, transitions at significance level 0.001, 0.039 and 0.014, respectively. On the other hand we have found that for Fe\,{\sc ii} 1608\,\AA\,\, transition distributions of H$_2$ candidate and DLA samples are not different at 0.94 significance level. 

The overall excess in C\,{\sc ii}, Si\,{\sc ii} and Al\,{\sc iii}
EWs of the H$_2$ candidate sample over the DLA sample can
possibly be interpreted as evidence for higher metal content.
In such an interpretation the corresponding similarity of the
Fe\,{\sc ii} 1608\,\AA\,\, EW distributions could reflect higher dust content
in the H$_2$ bearing systems. However, EWs are not
obviously related to column densities at low spectral resolution
and large EWs could just be a consequence of larger velocity
spread of the absorption.
Note however that there is a correlation between
velocity spread and metallicity \citep{Ledoux2006} which would
imply larger metallicities in the H$_2$ bearing candidates.

\section{Conclusion}
\label{conclusion}
\noindent

We have performed a systematic search for H$_2$ absorption in DLAs detected  towards quasars in the Sloan Digital Sky Survey releases 
DR7 and DR9. 
We have developed and used fully automatic procedures based on the profile fitting technique with modified $\chi^2$ function. 
The main difficulty of the search is the presence of the Ly-$\alpha$ forest which is dense at redshifts under consideration ($z>2.2$) and 
can effectively mimic H$_2$ absorption lines at the resolution and S/N of SDSS spectra. We carefully checked our searching procedure 
and found that it yields a completeness of $>99\%$ for $\log$N$_{\rm H2}$ $>$18.5. However the number of false detections is rather high.  For each of the candidates, the probability of false identification was calculated using two techniques.
The first technique employs Monte-Carlo simulations to estimate the probability of an accidental fit of an absorption system in the Ly$\alpha$ forest. In this technique we used repeated random shifts for each H$_2$ absorption line in the particular spectrum. The advantage of this technique is to give an estimate of the false identification probability in the particular spectrum (i.e. the particular 
realization of the Ly$\alpha$ forest). The second technique uses a control sample to estimate the false identification probability. 
Control sample consists of the non-BAL non-DLA quasars from DR9 SDSS catalog which guarantees the absence of H$_2$ absorption systems. The false identification probability can be calculated as the identification 
rate of H$_2$ absorption systems in the control sample. We additionally have performed simulations of mock SDSS quasar spectra which were 
used to check the procedures and to increase statistics. We have 
found that the false identification probability depends on the H$_2$ column density and spectral properties, mainly on S/N, $z_{\rm QSO}$, and 
the number  of H$_2$ absorption lines present in the spectrum. This method allows us to estimate the detection limit of H$_2$ absorption systems 
in SDSS spectra, i.e. the spectral properties for which the false identification probability is low and therefore the identification of the H$_2$ absorption 
system is robust. We found that in SDSS data a reasonable detection limit for H$_2$ column density, $\log$N$_{\rm H_2}$, is higher than 19.
We also derived that the upper limit on
the fraction of DLAs with H$_2$ systems is statistically equal to 7 and 3\% for, respectively, $\log$N$_{\rm H_2}>$19 and 19.5.
These are as upper limits because we tend to overestimate H$_2$ column densities.
The false identification probabilities calculated by the Monte-Carlo sampling and control sample techniques agree with each other
very well.

We have selected 23 candidates of H$_2$ absorption systems with high confidence level which are promising candidates for follow up 
high resolution studies. Taking into account the derived upper limit on the fraction of DLAs with saturated H$_2$ lines $\sim 7 \%$ it leads us to the conclusion that only less than 3\% of SDSS spectra are suitable for confident detection of H$_2$ absorption systems.

We studied the properties of these candidates, namely, color excess, extinction and equivalent widths of metal lines. 
There is a 1$\sigma$ $r-i$ color excess and no significant $A_V$ extinction
in quasar spectra with an H2 candidate compared to standard DLA bearing quasar spectra.
We find larger C\,{\sc ii}, Si\,{\sc ii}, and Al\,{\sc iii} equivalent widths (EW) in
the H$_2$ candidate sample over the DLA sample but no significant
difference for Fe\,{\sc ii} EWs. This is probably related to a larger
spread in velocity of the absorption lines in the H$_2$ bearing sample
and therefore possibly larger metallicities.

The selected candidates would increase by a factor of two the number of known H$_2$ absorption systems at high redshift. It is important to note, that we can 
confidently identify only saturated H$_2$ systems with high column densities, $\log$N$_{\rm H_2} > 19$. To date there are only a few such systems 
detected at high redshift. The selected candidates will undoubtedly allow us to gather a large sample of HD detections.
The relative abundance of HD/H$_2$ molecules which provides important clues on the chemistry and star-formation history and also can be used to estimate the isotopic D/H ratio and consequently $\Omega_b$ -- the density of baryonic matter in the Universe \citep{Ivanchik2010, Balashev2010}. In addition these systems are unique objects to search for CO molecule 
\citep{Srianand2008}. This molecule is suitable to estimate the physical conditions in interstellar clouds and the CMBR temperature at high redshift \citep{Noterdaeme2010, Noterdaeme2011}. We have selected four candidates 
with column densities $\log$N$_{\rm H_2} > 20$. Such systems have never been observed at high redshifts towards QSO sightlines and 
would provide an exclusive opportunity to study
translucent clouds in the interstellar medium at high redshift. 

\vspace{2mm}{\footnotesize {\rm Acknowledgments.}
We are very grateful the anonymous referee for the detailed and careful reading our manuscript and many useful comments. 
This work was partially supported by a State Program ``Leading Scientific Schools of 
Russian Federation'' (grant NSh-294.2014.2). SB and SK partially supported by the RF Presedent Programme (grant MK-4861.2013.2).
\bibliographystyle{mn2e}
\bibliography{H2}

\begin{thebibliography}{53}
\expandafter\ifx\csname natexlab\endcsname\relax\def\natexlab#1{#1}\fi

\bibitem[{{Abazajian} {et~al}\mbox{.}(2009){Abazajian}, {Adelman-McCarthy},
  {Ag{\"u}eros}, {Allam}, {Allende Prieto}, {An}, {Anderson}, {Anderson},
  {Annis}, {Bahcall}, \& et~al.}]{Abazajian2009}
{Abazajian} K.~N. {et~al.}, 2009, \apjs, 182, 543

\bibitem[{{Ahn} {et~al}\mbox{.}(2012){Ahn}, {Alexandroff}, {Allende Prieto},
  {Anderson}, {Anderton}, {Andrews}, {Aubourg}, {Bailey}, {Balbinot}, {Barnes},
  \& et~al.}]{Ahn2012}
{Ahn} C.~P. {et~al.}, 2012, \apjs, 203, 21

\bibitem[{{Balashev} {et~al}\mbox{.}(2010){Balashev}, {Ivanchik}, \&
  {Varshalovich}}]{Balashev2010}
{Balashev} S.~A., {Ivanchik} A.~V., {Varshalovich} D.~A., 2010, Astronomy
  Letters, 36, 761

\bibitem[{{Busca} {et~al}\mbox{.}(2013){Busca}, {Delubac}, {Rich}, {Bailey},
  {Font-Ribera}, {Kirkby}, {Le Goff}, {Pieri}, {Slosar}, {Aubourg}, {Bautista},
  {Bizyaev}, {Blomqvist}, {Bolton}, {Bovy}, {Brewington}, {Borde}, {Brinkmann},
  {Carithers}, {Croft}, {Dawson}, {Ebelke}, {Eisenstein}, {Hamilton}, {Ho},
  {Hogg}, {Honscheid}, {Lee}, {Lundgren}, {Malanushenko}, {Malanushenko},
  {Margala}, {Maraston}, {Mehta}, {Miralda-Escud{\'e}}, {Myers}, {Nichol},
  {Noterdaeme}, {Olmstead}, {Oravetz}, {Palanque-Delabrouille}, {Pan},
  {P{\^a}ris}, {Percival}, {Petitjean}, {Roe}, {Rollinde}, {Ross}, {Rossi},
  {Schlegel}, {Schneider}, {Shelden}, {Sheldon}, {Simmons}, {Snedden},
  {Tinker}, {Viel}, {Weaver}, {Weinberg}, {White}, {Y{\`e}che}, \&
  {York}}]{Busca2013}
{Busca} N.~G. {et~al.}, 2013, \aap, 552, A96

\bibitem[{{Crighton} {et~al}\mbox{.}(2013){Crighton}, {Bechtold}, {Carswell},
  {Dav{\'e}}, {Foltz}, {Jannuzi}, {Morris}, {O'Meara}, {Prochaska}, {Schaye},
  \& {Tejos}}]{Crighton2013}
{Crighton} N.~H.~M. {et~al.}, 2013, \mnras, 433, 178

\bibitem[{{Cui} {et~al}\mbox{.}(2005){Cui}, {Bechtold}, {Ge}, \&
  {Meyer}}]{Cui2005}
{Cui} J., {Bechtold} J., {Ge} J., {Meyer} D.~M., 2005, \apj, 633, 649

\bibitem[{{Ellison} {et~al}\mbox{.}(2001){Ellison}, {Yan}, {Hook}, {Pettini},
  {Wall}, \& {Shaver}}]{Ellison2001}
{Ellison} S.~L., {Yan} L., {Hook} I.~M., {Pettini} M., {Wall} J.~V., {Shaver}
  P., 2001, \aap, 379, 393

\bibitem[{{Fynbo} {et~al}\mbox{.}(2011){Fynbo}, {Ledoux}, {Noterdaeme},
  {Christensen}, {M{\o}ller}, {Durgapal}, {Goldoni}, {Kaper}, {Krogager},
  {Laursen}, {Maund}, {Milvang-Jensen}, {Okoshi}, {Rasmussen}, {Thorsen},
  {Toft}, \& {Zafar}}]{Fynbo2011}
{Fynbo} J.~P.~U. {et~al.}, 2011, \mnras, 413, 2481

\bibitem[{{Ge} \& {Bechtold}(1997)}]{Ge1997}
{Ge} J., {Bechtold} J., 1997, \apjl, 477, L73

\bibitem[{{Guimar{\~a}es} {et~al}\mbox{.}(2012){Guimar{\~a}es}, {Noterdaeme},
  {Petitjean}, {Ledoux}, {Srianand}, {L{\'o}pez}, \& {Rahmani}}]{Guimaraes2012}
{Guimar{\~a}es} R., {Noterdaeme} P., {Petitjean} P., {Ledoux} C., {Srianand}
  R., {L{\'o}pez} S., {Rahmani} H., 2012, \aj, 143, 147

\bibitem[{{Ivanchik} {et~al}\mbox{.}(2005){Ivanchik}, {Petitjean},
  {Varshalovich}, {Aracil}, {Srianand}, {Chand}, {Ledoux}, \&
  {Boiss{\'e}}}]{Ivanchik2005}
{Ivanchik} A., {Petitjean} P., {Varshalovich} D., {Aracil} B., {Srianand} R.,
  {Chand} H., {Ledoux} C., {Boiss{\'e}} P., 2005, \aap, 440, 45

\bibitem[{{Ivanchik} {et~al}\mbox{.}(2010){Ivanchik}, {Petitjean}, {Balashev},
  {Srianand}, {Varshalovich}, {Ledoux}, \& {Noterdaeme}}]{Ivanchik2010}
{Ivanchik} A.~V., {Petitjean} P., {Balashev} S.~A., {Srianand} R.,
  {Varshalovich} D.~A., {Ledoux} C., {Noterdaeme} P., 2010, \mnras, 404, 1583

\bibitem[{{Jorgenson} {et~al}\mbox{.}(2010){Jorgenson}, {Wolfe}, \&
  {Prochaska}}]{Jorgenson2010}
{Jorgenson} R.~A., {Wolfe} A.~M., {Prochaska} J.~X., 2010, \apj, 722, 460

\bibitem[{{Jorgenson} {et~al}\mbox{.}(2009){Jorgenson}, {Wolfe}, {Prochaska},
  \& {Carswell}}]{Jorgenson2009}
{Jorgenson} R.~A., {Wolfe} A.~M., {Prochaska} J.~X., {Carswell} R.~F., 2009,
  \apj, 704, 247

\bibitem[{{Jorgenson} {et~al}\mbox{.}(2006){Jorgenson}, {Wolfe}, {Prochaska},
  {Lu}, {Howk}, {Cooke}, {Gawiser}, \& {Gelino}}]{Jorgenson2006}
{Jorgenson} R.~A., {Wolfe} A.~M., {Prochaska} J.~X., {Lu} L., {Howk} J.~C.,
  {Cooke} J., {Gawiser} E., {Gelino} D.~M., 2006, \apj, 646, 730

\bibitem[{{Krogager} {et~al}\mbox{.}(2012){Krogager}, {Fynbo}, {M{\o}ller},
  {Ledoux}, {Noterdaeme}, {Christensen}, {Milvang-Jensen}, \&
  {Sparre}}]{Krogager2012}
{Krogager} J.-K., {Fynbo} J.~P.~U., {M{\o}ller} P., {Ledoux} C., {Noterdaeme}
  P., {Christensen} L., {Milvang-Jensen} B., {Sparre} M., 2012, \mnras, 424, L1

\bibitem[{{Kr{\"u}hler} {et~al}\mbox{.}(2013){Kr{\"u}hler}, {Ledoux}, {Fynbo},
  {Vreeswijk}, {Schmidl}, {Malesani}, {Christensen}, {De Cia}, {Hjorth},
  {Jakobsson}, {Kann}, {Kaper}, {Vergani}, {Afonso}, {Covino}, {de Ugarte
  Postigo}, {D'Elia}, {Filgas}, {Goldoni}, {Greiner}, {Hartoog},
  {Milvang-Jensen}, {Nardini}, {Piranomonte}, {Rossi},
  {S{\'a}nchez-Ram{\'{\i}}rez}, {Schady}, {Schulze}, {Sudilovsky}, {Tanvir},
  {Tagliaferri}, {Watson}, {Wiersema}, {Wijers}, \& {Xu}}]{Kruhler2013}
{Kr{\"u}hler} T. {et~al.}, 2013, ArXiv e-prints

\bibitem[{{Krumholz}(2012)}]{Krumholz2012}
{Krumholz} M.~R., 2012, \apj, 759, 9

\bibitem[{{Ledoux} {et~al}\mbox{.}(2003){Ledoux}, {Petitjean}, \&
  {Srianand}}]{Ledoux2003}
{Ledoux} C., {Petitjean} P., {Srianand} R., 2003, \mnras, 346, 209

\bibitem[{{Ledoux} {et~al}\mbox{.}(2006){Ledoux}, {Petitjean}, \&
  {Srianand}}]{Ledoux2006}
{Ledoux} C., {Petitjean} P., {Srianand} R., 2006, \apjl, 640, L25

\bibitem[{{Ledoux} {et~al}\mbox{.}(2002){Ledoux}, {Srianand}, \&
  {Petitjean}}]{Ledoux2002}
{Ledoux} C., {Srianand} R., {Petitjean} P., 2002, \aap, 392, 781

\bibitem[{{Levshakov} \& {Varshalovich}(1985)}]{Levshakov1985}
{Levshakov} S.~A., {Varshalovich} D.~A., 1985, \mnras, 212, 517

\bibitem[{{Malec} {et~al}\mbox{.}(2010){Malec}, {Buning}, {Murphy},
  {Milutinovic}, {Ellison}, {Prochaska}, {Kaper}, {Tumlinson}, {Carswell}, \&
  {Ubachs}}]{Malec2010}
{Malec} A.~L. {et~al.}, 2010, \mnras, 403, 1541

\bibitem[{{Meiksin}(2009)}]{Meiksin2009}
{Meiksin} A.~A., 2009, Reviews of Modern Physics, 81, 1405

\bibitem[{{Noterdaeme} {et~al}\mbox{.}(2008{\natexlab{a}}){Noterdaeme},
  {Ledoux}, {Petitjean}, \& {Srianand}}]{Noterdaeme2008}
{Noterdaeme} P., {Ledoux} C., {Petitjean} P., {Srianand} R.,
  2008{\natexlab{a}}, \aap, 481, 327

\bibitem[{{Noterdaeme} {et~al}\mbox{.}(2012){Noterdaeme}, {Petitjean},
  {Carithers}, {P{\^a}ris}, {Font-Ribera}, {Bailey}, {Aubourg}, {Bizyaev},
  {Ebelke}, {Finley}, {Ge}, {Malanushenko}, {Malanushenko},
  {Miralda-Escud{\'e}}, {Myers}, {Oravetz}, {Pan}, {Pieri}, {Ross},
  {Schneider}, {Simmons}, \& {York}}]{Noterdaeme2012}
{Noterdaeme} P. {et~al.}, 2012, \aap, 547, L1

\bibitem[{{Noterdaeme} {et~al}\mbox{.}(2010){Noterdaeme}, {Petitjean},
  {Ledoux}, {L{\'o}pez}, {Srianand}, \& {Vergani}}]{Noterdaeme2010}
{Noterdaeme} P., {Petitjean} P., {Ledoux} C., {L{\'o}pez} S., {Srianand} R.,
  {Vergani} S.~D., 2010, \aap, 523, A80

\bibitem[{{Noterdaeme} {et~al}\mbox{.}(2009){Noterdaeme}, {Petitjean},
  {Ledoux}, \& {Srianand}}]{Noterdaeme2009}
{Noterdaeme} P., {Petitjean} P., {Ledoux} C., {Srianand} R., 2009, \aap, 505,
  1087

\bibitem[{{Noterdaeme} {et~al}\mbox{.}(2008{\natexlab{b}}){Noterdaeme},
  {Petitjean}, {Ledoux}, {Srianand}, \& {Ivanchik}}]{Noterdaeme2008b}
{Noterdaeme} P., {Petitjean} P., {Ledoux} C., {Srianand} R., {Ivanchik} A.,
  2008{\natexlab{b}}, \aap, 491, 397

\bibitem[{{Noterdaeme} {et~al}\mbox{.}(2011){Noterdaeme}, {Petitjean},
  {Srianand}, {Ledoux}, \& {L{\'o}pez}}]{Noterdaeme2011}
{Noterdaeme} P., {Petitjean} P., {Srianand} R., {Ledoux} C., {L{\'o}pez} S.,
  2011, \aap, 526, L7

\bibitem[{{P{\^a}ris} {et~al}\mbox{.}(2012){P{\^a}ris}, {Petitjean}, {Aubourg},
  {Bailey}, {Ross}, {Myers}, {Strauss}, {Anderson}, {Arnau}, {Bautista},
  {Bizyaev}, {Bolton}, {Bovy}, {Brandt}, {Brewington}, {Browstein}, {Busca},
  {Capellupo}, {Carithers}, {Croft}, {Dawson}, {Delubac}, {Ebelke},
  {Eisenstein}, {Engelke}, {Fan}, {Filiz Ak}, {Finley}, {Font-Ribera}, {Ge},
  {Gibson}, {Hall}, {Hamann}, {Hennawi}, {Ho}, {Hogg}, {Ivezi{\'c}}, {Jiang},
  {Kimball}, {Kirkby}, {Kirkpatrick}, {Lee}, {Le Goff}, {Lundgren}, {MacLeod},
  {Malanushenko}, {Malanushenko}, {Maraston}, {McGreer}, {McMahon},
  {Miralda-Escud{\'e}}, {Muna}, {Noterdaeme}, {Oravetz},
  {Palanque-Delabrouille}, {Pan}, {Perez-Fournon}, {Pieri}, {Richards},
  {Rollinde}, {Sheldon}, {Schlegel}, {Schneider}, {Slosar}, {Shelden}, {Shen},
  {Simmons}, {Snedden}, {Suzuki}, {Tinker}, {Viel}, {Weaver}, {Weinberg},
  {White}, {Wood-Vasey}, \& {Y{\`e}che}}]{Paris2012}
{P{\^a}ris} I. {et~al.}, 2012, \aap, 548, A66

\bibitem[{{P{\^a}ris} {et~al}\mbox{.}(2011){P{\^a}ris}, {Petitjean},
  {Rollinde}, {Aubourg}, {Busca}, {Charlassier}, {Delubac}, {Hamilton}, {Le
  Goff}, {Palanque-Delabrouille}, {Peirani}, {Pichon}, {Rich},
  {Vargas-Maga{\~n}a}, \& {Y{\`e}che}}]{Paris2011}
{P{\^a}ris} I. {et~al.}, 2011, \aap, 530, A50

\bibitem[{{Petitjean} {et~al}\mbox{.}(2006){Petitjean}, {Ledoux}, {Noterdaeme},
  \& {Srianand}}]{Petitjean2006}
{Petitjean} P., {Ledoux} C., {Noterdaeme} P., {Srianand} R., 2006, \aap, 456,
  L9

\bibitem[{{Prochaska} {et~al}\mbox{.}(2009){Prochaska}, {Sheffer}, {Perley},
  {Bloom}, {Lopez}, {Dessauges-Zavadsky}, {Chen}, {Filippenko},
  {Ganeshalingam}, {Li}, {Miller}, \& {Starr}}]{Prochaska2009b}
{Prochaska} J.~X. {et~al.}, 2009, \apjl, 691, L27

\bibitem[{{Prochaska} \& {Wolfe}(2009)}]{Prochaska2009}
{Prochaska} J.~X., {Wolfe} A.~M., 2009, \apj, 696, 1543

\bibitem[{{Quider} {et~al}\mbox{.}(2011){Quider}, {Nestor}, {Turnshek}, {Rao},
  {Monier}, {Weyant}, \& {Busche}}]{Quider2011}
{Quider} A.~M., {Nestor} D.~B., {Turnshek} D.~A., {Rao} S.~M., {Monier} E.~M.,
  {Weyant} A.~N., {Busche} J.~R., 2011, \aj, 141, 137

\bibitem[{{Rachford} {et~al}\mbox{.}(2002){Rachford}, {Snow}, {Tumlinson},
  {Shull}, {Blair}, {Ferlet}, {Friedman}, {Gry}, {Jenkins}, {Morton}, {Savage},
  {Sonnentrucker}, {Vidal-Madjar}, {Welty}, \& {York}}]{Rachford2002}
{Rachford} B.~L. {et~al.}, 2002, \apj, 577, 221

\bibitem[{{Rahmani} {et~al}\mbox{.}(2013){Rahmani}, {Wendt}, {Srianand},
  {Noterdaeme}, {Petitjean}, {Molaro}, {Whitmore}, {Murphy}, {Centurion},
  {Fathivavsari}, {D'Odorico}, {Evans}, {Levshakov}, {Lopez}, {Martins},
  {Reimers}, \& {Vladilo}}]{Rahmani2013}
{Rahmani} H. {et~al.}, 2013, ArXiv e-prints

\bibitem[{{Reimers} {et~al}\mbox{.}(2003){Reimers}, {Baade}, {Quast}, \&
  {Levshakov}}]{Reimers2003}
{Reimers} D., {Baade} R., {Quast} R., {Levshakov} S.~A., 2003, \aap, 410, 785

\bibitem[{{Schlafly} \& {Finkbeiner}(2011)}]{Schlafly2011}
{Schlafly} E.~F., {Finkbeiner} D.~P., 2011, \apj, 737, 103

\bibitem[{{Schlegel} {et~al}\mbox{.}(1998){Schlegel}, {Finkbeiner}, \&
  {Davis}}]{Schlegel1998}
{Schlegel} D.~J., {Finkbeiner} D.~P., {Davis} M., 1998, \apj, 500, 525

\bibitem[{{Schneider} {et~al}\mbox{.}(2010){Schneider}, {Richards}, {Hall},
  {Strauss}, {Anderson}, {Boroson}, {Ross}, {Shen}, {Brandt}, {Fan}, {Inada},
  {Jester}, {Knapp}, {Krawczyk}, {Thakar}, {Vanden Berk}, {Voges}, {Yanny},
  {York}, {Bahcall}, {Bizyaev}, {Blanton}, {Brewington}, {Brinkmann},
  {Eisenstein}, {Frieman}, {Fukugita}, {Gray}, {Gunn}, {Hibon}, {Ivezi{\'c}},
  {Kent}, {Kron}, {Lee}, {Lupton}, {Malanushenko}, {Malanushenko}, {Oravetz},
  {Pan}, {Pier}, {Price}, {Saxe}, {Schlegel}, {Simmons}, {Snedden}, {SubbaRao},
  {Szalay}, \& {Weinberg}}]{Schneider2010}
{Schneider} D.~P. {et~al.}, 2010, \aj, 139, 2360

\bibitem[{{Smee} {et~al}\mbox{.}(2013){Smee}, {Gunn}, {Uomoto}, {Roe},
  {Schlegel}, {Rockosi}, {Carr}, {Leger}, {Dawson}, {Olmstead}, {Brinkmann},
  {Owen}, {Barkhouser}, {Honscheid}, {Harding}, {Long}, {Lupton}, {Loomis},
  {Anderson}, {Annis}, {Bernardi}, {Bhardwaj}, {Bizyaev}, {Bolton},
  {Brewington}, {Briggs}, {Burles}, {Burns}, {Castander}, {Connolly},
  {Davenport}, {Ebelke}, {Epps}, {Feldman}, {Friedman}, {Frieman}, {Heckman},
  {Hull}, {Knapp}, {Lawrence}, {Loveday}, {Mannery}, {Malanushenko},
  {Malanushenko}, {Merrelli}, {Muna}, {Newman}, {Nichol}, {Oravetz}, {Pan},
  {Pope}, {Ricketts}, {Shelden}, {Sandford}, {Siegmund}, {Simmons}, {Smith},
  {Snedden}, {Schneider}, {SubbaRao}, {Tremonti}, {Waddell}, \&
  {York}}]{Smee2013}
{Smee} S.~A. {et~al.}, 2013, \aj, 146, 32

\bibitem[{{Songaila} {et~al}\mbox{.}(1994){Songaila}, {Cowie}, {Vogt}, {Keane},
  {Wolfei}, {Hu}, {Oren}, {Tytleri}, \& {Lanzetta}}]{Songaila1994}
{Songaila} A. {et~al.}, 1994, \nat, 371, 43

\bibitem[{{Srianand} {et~al}\mbox{.}(2008){Srianand}, {Noterdaeme}, {Ledoux},
  \& {Petitjean}}]{Srianand2008}
{Srianand} R., {Noterdaeme} P., {Ledoux} C., {Petitjean} P., 2008, \aap, 482,
  L39

\bibitem[{{Srianand} {et~al}\mbox{.}(2000){Srianand}, {Petitjean}, \&
  {Ledoux}}]{Srianand2000}
{Srianand} R., {Petitjean} P., {Ledoux} C., 2000, \nat, 408, 931

\bibitem[{{Srianand} {et~al}\mbox{.}(2005){Srianand}, {Petitjean}, {Ledoux},
  {Ferland}, \& {Shaw}}]{Srianand2005}
{Srianand} R., {Petitjean} P., {Ledoux} C., {Ferland} G., {Shaw} G., 2005,
  \mnras, 362, 549

\bibitem[{{Thompson}(1975)}]{Thompson1975}
{Thompson} R.~I., 1975, \aplett, 16, 3

\bibitem[{{Vanden Berk} {et~al}\mbox{.}(2001){Vanden Berk}, {Richards},
  {Bauer}, {Strauss}, {Schneider}, {Heckman}, {York}, {Hall}, {Fan}, {Knapp},
  {Anderson}, {Annis}, {Bahcall}, {Bernardi}, {Briggs}, {Brinkmann}, {Brunner},
  {Burles}, {Carey}, {Castander}, {Connolly}, {Crocker}, {Csabai}, {Doi},
  {Finkbeiner}, {Friedman}, {Frieman}, {Fukugita}, {Gunn}, {Hennessy},
  {Ivezi{\'c}}, {Kent}, {Kunszt}, {Lamb}, {Leger}, {Long}, {Loveday}, {Lupton},
  {Meiksin}, {Merelli}, {Munn}, {Newberg}, {Newcomb}, {Nichol}, {Owen}, {Pier},
  {Pope}, {Rockosi}, {Schlegel}, {Siegmund}, {Smee}, {Snir}, {Stoughton},
  {Stubbs}, {SubbaRao}, {Szalay}, {Szokoly}, {Tremonti}, {Uomoto}, {Waddell},
  {Yanny}, \& {Zheng}}]{VandenBerk2001}
{Vanden Berk} D.~E. {et~al.}, 2001, \aj, 122, 549

\bibitem[{{Varshalovich} {et~al}\mbox{.}(2001){Varshalovich}, {Ivanchik},
  {Petitjean}, {Srianand}, \& {Ledoux}}]{Varshalovich2001}
{Varshalovich} D.~A., {Ivanchik} A.~V., {Petitjean} P., {Srianand} R., {Ledoux}
  C., 2001, Astronomy Letters, 27, 683

\bibitem[{{Wendt} \& {Molaro}(2012)}]{Wendt2012}
{Wendt} M., {Molaro} P., 2012, \aap, 541, A69

\bibitem[{{York} {et~al}\mbox{.}(2000){York}, {Adelman}, {Anderson},
  {Anderson}, {Annis}, {Bahcall}, {Bakken}, {Barkhouser}, {Bastian}, {Berman},
  {Boroski}, {Bracker}, {Briegel}, {Briggs}, {Brinkmann}, {Brunner}, {Burles},
  {Carey}, {Carr}, {Castander}, {Chen}, {Colestock}, {Connolly}, {Crocker},
  {Csabai}, {Czarapata}, {Davis}, {Doi}, {Dombeck}, {Eisenstein}, {Ellman},
  {Elms}, {Evans}, {Fan}, {Federwitz}, {Fiscelli}, {Friedman}, {Frieman},
  {Fukugita}, {Gillespie}, {Gunn}, {Gurbani}, {de Haas}, {Haldeman}, {Harris},
  {Hayes}, {Heckman}, {Hennessy}, {Hindsley}, {Holm}, {Holmgren}, {Huang},
  {Hull}, {Husby}, {Ichikawa}, {Ichikawa}, {Ivezi{\'c}}, {Kent}, {Kim},
  {Kinney}, {Klaene}, {Kleinman}, {Kleinman}, {Knapp}, {Korienek}, {Kron},
  {Kunszt}, {Lamb}, {Lee}, {Leger}, {Limmongkol}, {Lindenmeyer}, {Long},
  {Loomis}, {Loveday}, {Lucinio}, {Lupton}, {MacKinnon}, {Mannery}, {Mantsch},
  {Margon}, {McGehee}, {McKay}, {Meiksin}, {Merelli}, {Monet}, {Munn},
  {Narayanan}, {Nash}, {Neilsen}, {Neswold}, {Newberg}, {Nichol}, {Nicinski},
  {Nonino}, {Okada}, {Okamura}, {Ostriker}, {Owen}, {Pauls}, {Peoples},
  {Peterson}, {Petravick}, {Pier}, {Pope}, {Pordes}, {Prosapio},
  {Rechenmacher}, {Quinn}, {Richards}, {Richmond}, {Rivetta}, {Rockosi},
  {Ruthmansdorfer}, {Sandford}, {Schlegel}, {Schneider}, {Sekiguchi}, {Sergey},
  {Shimasaku}, {Siegmund}, {Smee}, {Smith}, {Snedden}, {Stone}, {Stoughton},
  {Strauss}, {Stubbs}, {SubbaRao}, {Szalay}, {Szapudi}, {Szokoly}, {Thakar},
  {Tremonti}, {Tucker}, {Uomoto}, {Vanden Berk}, {Vogeley}, {Waddell}, {Wang},
  {Watanabe}, {Weinberg}, {Yanny}, {Yasuda}, \& {SDSS
  Collaboration}}]{York2000}
{York} D.~G. {et~al.}, 2000, \aj, 120, 1579

\bibitem[{{Zhu} \& {M{\'e}nard}(2013)}]{Zhu2013}
{Zhu} G., {M{\'e}nard} B., 2013, \apj, 770, 130

\end{thebibliography}
\label{lastpage}

\bsp

\end{document}